\documentclass[12pt,english]{article}
\pdfoutput=1

\usepackage[authordate,
backend=biber,
doi=only,
isbn=false,
sorting=nyt,
maxcitenames=3,
minbibnames=7,
maxbibnames=7,
uniquename=false,
sortcites=true]{biblatex-chicago}

\bibliography{bib/references.bib}

\AtEveryBibitem{\clearlist{note}\clearlist{language}\clearlist{issn}} % clears issn
\AtEveryBibitem{%
	\ifentrytype{online}{%
		\clearfield{urlyear}
		\clearfield{urlmonth}
		\clearfield{urlday}
		\clearfield{note}
		\clearlist{language}
	}{%
		\clearfield{eprint}%
		\clearfield{urlyear}
		\clearfield{urlmonth}
		\clearfield{urlday}
		\clearfield{note}
		\clearlist{language}
	}
}%

\usepackage{CJKutf8}
\newcommand{\chinese}[1]{\begin{CJK}{UTF8}{bkai}#1\end{CJK}}
\usepackage{mathptmx}
\usepackage{microtype}
\usepackage{amsmath}
\usepackage{bbm}
\usepackage[utf8]{inputenc}
\usepackage[T1]{fontenc}
\usepackage{babel}
\usepackage{setspace}
\usepackage{graphicx}
\usepackage{booktabs,dcolumn}
\usepackage{array}
\usepackage{multirow}
\usepackage[usenames,dvipsnames,svgnames,table]{xcolor}
\usepackage[export]{adjustbox}[2011/08/13]
\usepackage{enumitem}
\usepackage[text={16cm,24cm}]{geometry}
\usepackage{ragged2e}
\usepackage{csquotes}
\usepackage{float}
\usepackage[nolists, nomarkers]{endfloat}
\usepackage{blindtext}
\usepackage{titling}

\usepackage[hang, flushmargin, bottom, symbol]{footmisc}
\usepackage[colorlinks=true, linkcolor=blue, citecolor=blue, plainpages=false, pdfpagelabels=true, urlcolor=blue]{hyperref}
\usepackage{subcaption}
\usepackage{caption}
\date{September 8, 2021}
\newcolumntype{L}[1]{>{\raggedright\let\newline\\\arraybackslash\hspace{0pt}}m{#1}}
\newcolumntype{C}[1]{>{\centering\let\newline\\\arraybackslash\hspace{0pt}}m{#1}}
\newcolumntype{R}[1]{>{\raggedleft\let\newline\\\arraybackslash\hspace{0pt}}m{#1}}

\newcommand{\sym}[1]{\ifmmode^{#1}\else\(^{#1}\)\fi}

\usepackage{footnotebackref}
\newcommand{\PAPERJEL}{\textbf{JEL}: J15, I24, O15}

\newcommand{\PAPERKEYWORDS}{\textbf{Keywords}: school closure and consolidation, rural education, minoritized populations, China}

\newcommand{\PAPERKEYWORDSSEC}{
\section*{Keywords}
\PAPERKEYWORDS
}

\newcommand{\PAPERTITLE}{Fewer, better pathways for all?\\Intersectional impacts of rural school consolidation in China’s minority regions}

\newcommand{\AUTHORHANNUM}{Emily Hannum}
\newcommand{\AUTHORHANNUMURL}{https://orcid.org/0000-0003-2011-9984}
\newcommand{\AUTHORHANNUMINFO}{\href{\AUTHORHANNUMURL}{\AUTHORHANNUM}: Department of Sociology and Population Studies Center, University of Pennsylvania, 3718 Locust Walk, Philadelphia, PA 19104 (email:hannumem@sas.upenn.edu)}

\newcommand{\AUTHORWANG}{Fan Wang}
\newcommand{\AUTHORWANGURL}{https://orcid.org/0000-0003-2640-5420}
\newcommand{\AUTHORWANGINFO}{\href{\AUTHORWANGURL}{\AUTHORWANG}: Department of Economics, University of Houston, 3623 Cullen Boulevard, Houston, TX 77204 (email: fwang26@uh.edu)}

\newcommand{\ACKNOWLEDGMENTS}{
We gratefully acknowledge support from Penn's University Research Foundation and School of Arts and Sciences Research Opportunity Grant Programs and from Grand Challenges Canada (PI: Jere Behrman). We also acknowledge support from Scholar Grant GS040-A-18 from the Chiang Ching-Kuo Foundation for coverage of Wang's time. Thanks to Anita Lai for providing editorial suggestions. This paper has benefited from helpful comments provided by two anonymous reviewers and by seminar participants at the Department of Sociology at the Chinese University of Hong Kong, the Center for Social Research at Peking University, the Population Research Institute at Duke University, the Population Research Institute at Penn State University, and  the Education and Inequality Workshop at the University of Pennsylvania.  We also thank Jere Behrman, Aimee Chin, Elaine Liu, and Xiuqi Yang for suggestions and comments.}

\newcommand{\PAPERABSTRACT}{
Primary school consolidation---the closure of small community schools or their mergers into larger, better-resourced schools---is emerging as a significant policy response to changing demographics in middle income countries with large rural populations.  In China, large-scale consolidation took place in the early 21st century.  Because officially-recognized minority populations disproportionately reside in rural and remote areas, minority students were among those at elevated risk of experiencing school consolidation. We analyze heterogeneous effects of consolidation on educational attainment and reported national language ability in China by exploiting variations in closure timing across villages and cohorts captured in a 2011 survey of provinces and autonomous regions with substantial minority populations. We consider heterogeneous treatment effects across groups defined at the intersections of minority status, gender, and community ethnic composition and socioeconomic status.

Compared to villages with schools, villages whose schools had closed reported that the schools students now attended were better resourced, less likely to offer minority language of instruction, more likely to have Han teachers, farther away, and more likely to require boarding. Much more than Han youth, ethnic minority youth were negatively affected by closure, in terms of its impact on both educational attainment and written Mandarin facility. However, for both outcomes, significant penalties accruing to minority youth occurred only in the poorest villages. Penalties were generally heavier for girls, but in the most ethnically segregated minority villages, boys from minority families were highly vulnerable to closure effects on educational attainment and written Mandarin facility. Results show that intersections of minority status, gender, and community characteristics can delineate significant heterogeneities in policy impacts.\\
\PAPERJEL}

\newcommand{\HIGHLIGHT}{
\section*{Highlights}
\begin{enumerate}
    \item Primary school consolidation is a policy response to changing demographics in middle income countries with large rural populations.
    \item We analyze the effects of consolidation on educational attainment and language ability among rural minority populations in China.
    \item After closure, better resourced and farther away schools offer less minority language instructions and have more Han teachers.
    \item Minority youth were negatively affected by closure, in terms of both educational attainment and written Mandarin facility.
    \item Penalties were heavier for girls from poorer villages and for boys from more ethnically segregated minority villages.
\end{enumerate}
}

\newcommand{\PAPERDOIURL}{https://doi.org/10.1016/j.worlddev.2021.105734}
\newcommand{\PAPERINFO}{
This paper is published as: Hannum, Emily, and Fan Wang. ``Fewer, Better Pathways for All? Intersectional Impacts of Rural School Consolidation in China’s Minority Regions.'' World Development 151 (March 1, 2022): 105734. \url{\PAPERDOIURL}.
}

\begin{document}

%%%%%%%%%%%%%%%%%%%%%%%%%%%%%%%%%%%%%%%%
% Part I. Title Page, Authors, Abstract, etc. 
%%%%%%%%%%%%%%%%%%%%%%%%%%%%%%%%%%%%%%%%
\title{
\vspace*{-4em}
\singlespacing
\PAPERTITLE\thanks{\PAPERINFO}}

\author{
\AUTHORHANNUM{} and \AUTHORWANG{}\thanks{
\AUTHORHANNUMINFO;
\AUTHORWANGINFO.
\ACKNOWLEDGMENTS}}

\date{
\vspace*{-1.0em}
March 1, 2022}
\maketitle

\vspace*{-2em}
\begin{abstract}
\vspace*{-1.0em}
\singlespacing
\PAPERABSTRACT
\end{abstract}
\thispagestyle{empty}
\clearpage

\doublespacing
\PAPERKEYWORDSSEC
\HIGHLIGHT
\thispagestyle{empty}
\clearpage

%%%%%%%%%%%%%%%%%%%%%%%%%%%%%%%%%%%%
% Part II. Manuscript main text
%%%%%%%%%%%%%%%%%%%%%%%%%%%%%%%%%%%%
\pagenumbering{arabic}
\setcounter{page}{1}
\renewcommand*{\thefootnote}{\arabic{footnote}}
% Manuscript main text
%%%%%%%%%%%%%%%%%%%%%%%%%%%%%%%%%%%
% Main Text
%%%%%%%%%%%%%%%%%%%%%%%%%%%%%%%%%%%
\section{Introduction}
\label{subsec:intro}
School consolidation---the closure and merger of small schools to form larger schools---is a common strategy to economize and create a better-resourced environment for staff and students in the context of demographic changes \autocite[56]{echazarra_learning_2019}. In OECD countries, for example, a recent report observes that ``shrinking student numbers, limited access to qualified and experienced teachers and a relatively high proportion of disadvantaged students make the efficient provision of high-quality education in rural areas a difficult undertaking" and that ``consolidation may be unavoidable to guarantee adequate learning environments for all students'' \autocite[14, 129]{oecd_responsive_2018}. Consolidation is also emerging in some middle income countries with large rural populations,\footnote{For example, see press reports about consolidation such as \textcites{malik_apprehensions_2013, chowdhury_cramped_2017, saengpassa_ministrys_2017, harun_schools_2017, tawie_sarawak_2017}.} as fertility trends, population aging, and rural outmigration set the stage for demographic decline.\footnote{The top 10 countries in the world in terms of projected rural population decline between 2018 and 2050, in descending order, are China, India, Indonesia, Bangladesh, United States of America, Russian Federation, Thailand, Viet Nam, Brazil, and Iran \autocite[4, 46]{statista_regional_2020}.}  In these contexts, consolidation is often linked to concerns about performance, capacity to provide quality schooling at reasonable cost, and  “demographic scarcity” in the school-aged population.\footnote{See \textcite[923-925]{Tieken_Auldridge-Reveles_2019} for similar rationales for consolidation in urban and rural areas in the United States.}

Yet, a recent OECD report on learning in rural schools by \textcite[56]{echazarra_learning_2019} also observes that there are ``serious risks and costs associated with school consolidation, including ``increased transportation costs for students’ families and relocated staff, reduced parental involvement, deteriorated local services, reduced school choice and other social costs." The review also raises the concern that socio-economically
disadvantaged students are more negatively affected following school closure \autocite[56]{echazarra_learning_2019}. Whether consolidation enhances or detracts from educational opportunity for students from economically and socially marginalized populations is particularly important to address.\footnote{See \textcite{Tieken_Auldridge-Reveles_2019} for a discussion of evidence regarding the presence or absence of racial unevenness in the experience of school consolidation in United States.}

Extremely rapid fertility decline and urbanization  have accelerated pressures for system consolidation in China, and the policy response has been striking in pace and scale. In 2001, the State Council initiated a massive national push to consolidate educational infrastructure. The school consolidation initiative, implemented by county level officials, sought to address sparse demand, inefficiencies in provision, and perceived quality problems in rural education \autocite{mei_school_2015}. \textcite[141-143]{mei_school_2015} cite official statistics showing that between 1998 and 2010, the number of primary schools dropped by 58 percent.

For students in affected communities, consolidation often means traveling longer distances but attending better-equipped schools, though larger schools can also mean larger class sizes.  Distance can sometimes bring greater expense, particularly if children need to board. Studies published on the phenomenon as recently as 2014 and 2015 commented on the surprisingly paucity of empirical research on the consequences of this important policy \autocite{chen_poor_2014,mei_school_2015}. While a handful of empirical studies in English have begun to address implications of system consolidation for access, cost, and performance (for example, see \textcite{chan_harrell_2009, liu_effect_2010, dong_empirical_2013, chen_poor_2014, liu_qualityquantity_2014,  cai_has_2017,  hannum_estimating_2020, Hou_Wu_Song_2020}), the implications of system consolidation for patterns of stratification have not been fully explored.

In particular, how the policy has affected China's minority youth is a significant question. As of the 2015 intercensal survey, more than one in ten youth ages 10 to 19 in China were members of officially-recognized minority groups \autocite[3]{UNICEF_China_UNFPA_China_2018}.  Despite absolute gains in access over time, minority youth in China, as a group, remain educationally vulnerable. Focusing on compulsory education as a minimum standard for access, in 2015, the percent of adolescents aged 10 to 19 not on track to receive the nine-year cycle of compulsory education was fairly negligible among majority Han adolescents, at 1.7 percent, but was 7.2 percent among minority youth \autocite[7]{UNICEF_China_UNFPA_China_2018}.\footnote{Not on track includes youth who had never been to school, had graduated from primary school only, or had dropped out of primary school or junior secondary school.}

As a group, minority youth are more likely than others to attend poorly resourced schools \autocite[519]{gustafsson_mapping_2015}.  The access vulnerability of minority youth is closely tied to geographic marginalization \autocite[321-322]{chen_chapter_2016}.\footnote{\textcite{pan_birth_2021} suggest that an additional factor in differential educational outcomes is a legacy of the more stringent fertility controls imposed on Han families, which translated to smaller sibships, on average, for Han students.} Children who are members of minority populations, as a group, are more likely than members of the Han majority to reside in remote, poor, and rural areas.  In the 2010 census, 71.4 percent of ethnic minority children ages 0 to 17 lived in rural areas, compared with only 53.4 percent of Han children \autocite[2]{united_nations_childrens_fund_china_ethnic_2016}. A more recent estimate based on the 2015 intercensal survey indicated that nearly half of all ethnic minority adolescents aged 10 to 19—45.4 percent—lived in poor rural areas, counties officially-recognized at that time as poverty-stricken, compared to 24 percent of the total adolescent population \autocite[4]{UNICEF_China_UNFPA_China_2018}.\footnote{A substantial body of literature has emerged that investigates links among economic development, geography, and ethnic disparities in education and other outcomes (for reviews of literature on ethnicity and education in China, see \textcite{chen_chapter_2016} and \textcite{stevens_china_2014}.  For near-national evidence on educational testing differences by minority status, albeit evidence that does not include the experiences of youth in Xinjiang, Tibet,
Qinghai, Inner Mongolia, Ningxia, or Hainan, see \textcite{wang_math_2018}.  Speaking to the importance of an intersectional perspective, the minority-majority performance gap found by \textcite{wang_math_2018} is reversed in the urban and rural samples, though findings for the urban sample are not statistically significant.}  Thus, minority youth disproportionately reside in the kinds of communities that are appealing targets for school consolidation.

Beyond the general risks and benefits associated with consolidation, for minoritized communities, there may be additional considerations. For example, consolidated schools may offer better national language instruction, but could be less equipped to offer mother-tongue instruction, if children from multiple language groups are coming together in consolidated schools.\footnote{\textcite[493]{roche_articulating_2019} points out that while there are 56 recognized ethnic groups in China, an important, regularly updated reference work on global linguistic diversity registers many more languages---306 in the 2021 edition \autocite{eberhard_china_2021}. As minority group boundaries were defined and languages recognized, linguistic diversity within groups was downplayed as standards were chosen, which created challenges to full preservation of  non-standard languages spoken by some included within defined ethnic categories \autocites{harrell_linguistics_1993,dwyer_texture_1998, roche_articulating_2019}.  For example,  \textcite[494]{roche_articulating_2019} observes that Mongolians speak at least six distinct languages. Chakhar Mongolian is taught as standard in the Inner Mongolia Autonomous Region and is replacing not only
several languages spoken by Mongolians but also those of other ethnic
groups, such as the Evenki, even as a language shift from Mongolian to Mandarin is also underway. Roche describes similar phenomena of standard versions of Yi and Tibetan languages edging out languages spoken by some members of these groups.  Similarly, some Chinese topolects---not recognized as separate languages---are mutually unintelligible with Mandarin and are as distant from each other as Romance languages \autocites[81]{dwyer_texture_1998}{}. This linguistic complexity highlights that whether or not children are receiving mother tongue education is hard to fully parse with existing data.}  Facility with the national language might be viewed as essential for economic mobility by some rural minority parents, and if this were the case, the greater Mandarin focus in consolidated schools might appeal.  At the same time, a more Mandarin-focused teaching environment might create barriers to engagement and learning for some minority youth. A related possibility is that minority parents’ decisions about sending children to school are more sensitive to increased distance (signifying cost or safety risk) or less sensitive to the kinds of school resources on offer in consolidated schools. Families may be less inclined to allow their children to board at school if there are linguistic mismatches or contexts of ethnic tension. Linguistic minority parents may hesitate to send young children far away to board at a school if they cannot communicate with teachers. Linguistic barriers in centralized schools might thus counteract, to some degree, benefits of better facilities at these larger schools for minority youth.

Drawing on a 2011 rural household and community survey of provinces and autonomous regions with substantial minority populations, we analyze the impact of school consolidation on minority and Han youth in provinces and autonomous regions with substantial minority populations. The dataset contains both villages that experienced closure between 2000 and 2011 as well as those that did not. In closure villages, older individuals who were beyond primary school age at the village-specific year of closure were not directly impacted, but younger individuals had to transition from village schools to consolidated schools. Using a difference-in-differences strategy, we identify causal effects of school closure by exploiting the variations in closure and closure timing across villages, as well as variations in age at closure within closure villages.

In our empirical analysis, we first compare school features in consolidation villages and those that have maintained village schools. Next, we test whether ethnic minority youth were disproportionately affected when they experienced school closure.  We estimate impacts on educational attainment and reported facility in Mandarin, because consolidated schools are likely to promote the use of the national language from an earlier age. Importantly, given literature suggesting that gender, impoverishment, and the isolation of minoritized communities might condition impacts, we allow for heterogeneous treatment effects across groups defined at the intersections of minority status, gender, and community ethnic composition and socioeconomic status.

We show that average negative effects of closure on educational attainment and reported written Mandarin ability for minority youth mask considerable heterogeneities across groups defined by intersections with gender, community economic status, and community ethnic composition. Penalties for minority youth were pronounced in poorer villages, while there was no evidence of a significant minority penalty in wealthier villages. Penalties were generally heavier for minority girls. However, in the most ethnically segregated minority villages, boys from minority families were highly vulnerable to closure effects on educational attainment and written Mandarin facility.

The remainder of this paper is organized in the following sections. Section 1 provides background on school consolidation and stratification, in comparative perspective and in China. In Section 2, we describe the data and estimation strategy. Section 3 presents and discusses estimation results. Section 4 concludes.
\section{Motivation and research questions}

\subsection{Consolidation in global perspective}
\label{System consolidation and educational stratification in comparative perspective}

\subsubsection{Global trends}
 In recent years, school consolidation has been advanced as a strategy to reduce costs and improve educational quality in the context of new demographic realities in many countries, with the idea being that consolidated schools have the potential to provide better facilities and more academic and extra-curricular programs and activities in an economical fashion \autocite[56]{echazarra_learning_2019}.
 For example, in a majority of OECD countries, only 20 percent or less of children up to the age of 14 live in
rural areas--a share which is even lower in the 15-29 age group--and this situation is creating what a recent report characterized as ``unsustainable" excess
capacities in rural areas \autocite[125]{oecd_responsive_2018}.
The same report observes that consolidation has often been proposed to ``increase the size, improve the resources and broaden the educational [offerings] of the remaining rural
schools" \autocite[128]{oecd_responsive_2018}.

The strategy is emerging as a response to changing demographics in some middle income countries with large rural populations. For example, in the case of Brazil, official education statistics show that the number of rural primary schools dropped 31 percent between 2007 and 2017, from 88,386 rural primary schools to 60,694 \autocite{brazil_ministry_of_education_sinopses_2020}.  Thailand's Ministry of Education recently announced a plan to merge thousands of small schools with fewer than 120 students each with other schools within a six-kilometer radius \autocite{saengpassa_ministrys_2017}. In Punjab, Pakistan, new reports indicate that of 51,602 primary and middle public schools in the province, about 5,500 primary schools have been merged \autocite{malik_apprehensions_2013}; in Rajasthan, India, in 2014, the government merged 17,000 of the over 80,000 government schools in the state with other schools, with more mergers planned \autocite{chowdhury_cramped_2017}. Further, India's National Education Policy 2020 recommends consolidation of schools (section 7.4) and the creation of `school complexes/clusters' (section 7.6) to address problems of small rural schools \autocites[3]{ramanand_national_2020}{india_ministry_of_human_resource_development_national_2021}.

\subsubsection{Consolidation and educational stratification}

The implications of rural school consolidation for educational attainment and educational stratification in low- and middle-income countries are not yet well theorized or researched.\footnote{See \textcite{Tieken_Auldridge-Reveles_2019} and \textcite{echazarra_learning_2019}, respectively, for reviews of research in the US and OECD countries.} When considering the impact of consolidations on inequality, one important question is, which communities experience closures? In principle, there are three common, related rationales for school closure decisions: demographic, economic, and performance-based. Demographic scarcity–-too few students for a school---is a common factor cited in closure decisions. Economic efficiency is also commonly cited–-there may be insufficient resources to provide curricular offerings in small, sparsely-populated schools. Finally, schools whose students perform poorly are also commonly targeted, with the hope that the better resources available in a consolidated school will address performance problems. These conditions are often associated with  socioeconomically marginalized communities, whether poorer urban communities or remote rural ones. As \textcite{Caven_2019} has observed in a very different setting---a case study of school closures in the city of Philadelphia---the metrics involved in selecting schools for closure tend to predispose poor and minority communities to institutional loss.\footnote{For a similar argument made more broadly about the United States, see \textcite{Tieken_Auldridge-Reveles_2019}.}

Separate from the question of whether there are disparate exposures to school consolidation, and more directly tied to the purpose of this paper, is whether school consolidation may have heterogeneous impacts on different groups of children. There is not yet a great deal of evidence on this point.  In one of the first evaluations of the policy in India, \textcite{bordoloi_school_2019} investigated the impact of closures in Rajasthan. Basic infrastructure was better in consolidated schools, though the number of students per classroom did increase post consolidation \autocite[22]{bordoloi_school_2019}. However, \textcite[23]{bordoloi_school_2019} report a greater decline in enrollment in consolidated schools compared to all government schools across Rajasthan, and that impacts were not equally borne. The authors conclude, ``worryingly, the decline in [enrollment] seems to be the highest for students with disability, followed by that of [Scheduled Caste] and [Scheduled Tribe] students" \autocite[23]{bordoloi_school_2019}.\footnote{\textcite[5]{ramanand_national_2020} cites a Save the Children report \autocite{rao2017school} about school mergers in Telangana, Odisha, and Rajasthan, which found that enrollments were negatively affected by school mergers, and that ``harm was more serious for students belonging to marginalized communities like Dalits and Tribes." The study reported no evidence of school mergers improving the quality of learning; instead, the authors reported inadequate numbers of teachers in cramped classrooms.}

One common effect of consolidation is to increase distance to schools, and the effects of distance might be heterogeneous.  A cross-national analysis of Demographic and Health Surveys carried out in the 1990s considered the relationship between the school enrollment of 6 to 14 year-olds and the distance to primary and secondary schools in 21 rural areas of low-income countries \autocite{Filmer_2007}. Findings suggested that distance and enrollment are statistically significantly related, but magnitudes of the associations are small: simulating large reductions in distance yielded only small increases in average school participation \autocite{Filmer_2007}.  A multi-level analysis of survey data from 220,000 children in 340 districts of 30 developing countries estimated that parental decisions regarding children's enrollment were associated with distance from school, after adjusting for other school, family and community characteristics \autocite{huisman_effects_2009}. One recent review suggested that greater distance to school or the absence of a nearby school carries stronger negative implications for girls’ than boys’ enrollments in many settings \autocite[25]{king_todays_2015}.  Similarly, \textcite[1632]{Glick_2008} reports that many studies have suggested that girls’ schooling responds more strongly than boys’ to changes in school distance or availability, while the opposite is very rarely found. This pattern could emerge if families are more worried about girls’ safety than boys', if parents depend more on girls' labor at home than boys', if parents are less willing to pay added expenses for girls than boys, or if parents respond differently to the improved resources in consolidated schools for girls than for boys.\footnote{See \textcite [77-78]{Jayachandran_2015} for a brief discussion of distance and other concerns about girls' enrollment).} However, \textcite{Glick_2008} also acknowledges that greater vulnerability to distance for girls is not a pattern that is consistent in cross-national studies.  \textcite{Filmer_2007}'s cross-national analysis of Demographic and Health Surveys found little impact of increasing the availability of primary schools on gender inequality.   \textcite[923]{Filmer_2007} plotted national male-female ratios of enrollment from a simulation of the impact of reducing the distance to all primary schools to zero against the actual ratio of male to female enrollment and found that almost all points fell on the 45 degree line. Along the same lines, \textcite{huisman_effects_2009} found similar magnitudes of effect of distance on enrollment for girls and boys at ages 8 to 11.

\subsubsection{Consolidation and linguistic and cultural minority populations}

In societies with significant linguistic and cultural minority populations, it is possible that consolidation could carry different implications for minoritized children than for other children. For example, the educational benefits of improved resources might be tempered if accompanied by reduced access to mother-tongue education. To the extent that consolidation implies switching from a small village school offering mother-tongue language of instruction to a larger, possibly multi-ethnic consolidated school offering national language of instruction, the shift could present significant barriers to learning for children. A number of studies have sought to identify the impact of mother-tongue education on educational and economic outcomes.  In India, \textcite[473]{Jain_2017} compared districts in which the official language matched the district’s language and some where it did not. Linguistically mismatched districts had 18.8 percent lower literacy rates and 27.6 percent lower college graduation rates, which the authors attribute to difficulty in acquiring education due to a different medium of instruction in schools. Further, \textcite{Jain_2017} found that educational achievement recovered in mismatched districts after a  1956 reorganization of Indian states on linguistic lines.

In South Africa, \textcite{Eriksson_2014} examined the impact of the 1955 Bantu Education Act, which mandated eight years of mother-tongue education for Black students. \textcite[311-313]{Eriksson_2014} drew on the 1980 South African census and estimated a difference-in-differences model to reveal the effect of increasing potential exposure to mother-tongue instruction for Black students from four to six years. Results show positive effects on wages. Evidence also suggested that English speaking ability \textit{increased} as a result of the policy, but only in predominantly English parts of the country. Finally, the paper found larger effects on wages and educational attainment for women than men, consistent with the idea that mother-tongue education increased accessibility more for girls than boys.

However, a review of these and other pieces by  \textcite{Ginsburgh_Weber_2020} highlights the mixed and context-specific evidence about implications of mother-tongue instruction, versus national, colonial, or international language instruction, for schooling and returns to schooling. A significant counterexample to the importance of mother-tongue education is provided in \textcite{Angrist_Chin_Godoy_2008}. The authors argue, with evidence from Puerto Rico during a period of change from English to Spanish language of instruction, that changes in language of instruction were not significantly associated with changes in reported English proficiency.

Apart from considerations of learning, it is conceivable that the prospects presented by consolidated schools might be weighed differently by families from linguistic and cultural minorities.  Differences might emerge if minority parents, more than others, had concerns about risk in a larger school, about children's ability to communicate with teachers, or about cultural preservation.  Differences could also emerge if minority parents perceived fewer prospects for their children realizing economic returns from better-resourced, consolidated schools.

\subsubsection{Intersectional perspectives}

Theoretically, the impact of consolidation in the case of minoritized youth might also be tied to other domains of social status. An intersectional perspective suggests that ``individuals are simultaneously situated in multiple social structures, which interact in complex ways to influence their experiences, social relations, and well being" \autocite[540]{tas_gender_2014}.  With regard to education, evidence from some countries suggests that significant cumulative disadvantages in education can occur for children at the nexus of female status and minority or indigenous status \autocite{tas_gender_2014}, perhaps especially under conditions of material deprivation. However, we have not found studies of consolidation that adopt an intersectional perspective.

In short, consolidation might bring particular educational benefits to children from minority communities, if they have full access to better-resourced schools. However, benefits could be tempered by the potential educational costs of linguistic mismatch or if the risk associated with children leaving home communities, to larger schools, is perceived to be greater by parents in minority communities.  The contrast between consolidated and village schools--both in terms of resources and in terms of language and cultural difference--might differ for boys and girls, and might be most pronounced in economically marginalized communities or more isolated minority communities.
 \subsection{Consolidation and stratification in China}
\label{Consolidation and stratification in China}
\subsubsection{Consolidation policies}
National school consolidation policies in China commenced in 2001, though \textcite{dai_cost_2017} report that consolidation experiments were piloted in some provinces in 1993.
The State Council's ``Decision on Basic Education Reform and Development'' \autocite{state_council_state_2001} required local governments to make reasonable adjustments to schools' geographic distribution to improve efficiency. As elsewhere, concerns with quality and efficiency loomed large  \autocite[see, for example,][]{liu_closures_2013, xie_consolidating_2013}. Resource-constrained counties may have faced particular pressures toward consolidation \autocite{fan_reasons_2013}.
By 2012, the Ministry of Education and then the General Office of the State Council issued documents calling for an end to consolidation \autocite{state_council_2012}.

However, persistent population decline in rural China continues to exert demographic pressures toward consolidation.  From a long-term perspective, China ranks at the top of all countries in terms of projected rural population loss in the coming years: the rural population of China is projected to be around 305 million less in 2050 than it was in 2018 \autocite[4,46]{statista_regional_2020}. More immediately, \textcite[93]{Hou_Wu_Song_2020} report that ongoing fiscal pressures create incentives for local governments to encourage the practice.  Policy debates continue, given concerns that have arisen around consolidated schools. A recent communique from the State Council committed to support for the establishment of small-scale rural schools (primary schools in villages and teaching points of less than 100 students) and improvement of township boarding schools \autocite{state_council_guiding_2018}; see discussion of this communique in \textcite[108]{NWCCW_UNICEF_NBS_2018}.

\subsubsection{Impact of school consolidation}
Several studies have investigated the impact of consolidation on student achievement \autocite{liu_effect_2010, mo_transfer_2012, chen_poor_2014}. \textcite{liu_effect_2010} find that the timing of consolidation in students' lives mattered: higher-grade students' grades rose after merging, while grades of younger students fell. \textcite{mo_transfer_2012} and \textcite{chen_poor_2014} find that elementary school students' academic performance improved when they transferred from less centralized schools to more-centralized schools.  \label{rr:citechen} Boarding at primary school can become necessary with longer distances that come with consolidation.  However, as discussed in \textcite{chen_poor_2014}, the need to board at school at early ages may jeopardize the benefits of centralized schools. Parents may not wish to avail themselves of  centralized schools if these schools are too far for daily commuting.

Are certain groups more vulnerable than others to school closure? Some studies have suggested that economically disadvantaged families may be most vulnerable due to costs imposed by consolidation.  \textcite{cai_has_2017} found that the compulsory school consolidation program increased educational expenditures, including expenditures on transportation and boarding due to greater distance to school. \textcite{zhao_increasingly_2015} found that children from poorer families had  difficulties paying for a bus or boarding at school and are more likely to endure longer commutes. Other studies suggest that some poor rural families might be more likely to shoulder costs for boys than girls: research in China suggests that girls' educational attainment has been more susceptible than boys' to poverty  \autocite{cherng2013community, liu2017early}.  Earlier research has also shown that rural girls are particularly affected by school consolidation, possibly because enrollment decisions for boys are more sensitive to quality, while those for girls are more sensitive to distance \autocite{hannum_estimating_2020}.

Another study has directly investigated socioeconomic and gender differences in the implications of consolidation.  \textcite{Hou_Wu_Song_2020} analyze survey data from 137 township schools with boarding facilities to study ``premature" early grade boarding that resulted from consolidation, and its impact on human capital accumulation.  They show that children from more adverse family circumstances are at greater risk than others to be early-grade boarders. Families were more likely to send boys than  girls to be early-grade boarders. Further analyses showed that early-grade boarding was associated with lower test scores, more depression, and bullying victimization.  Subsample analyses showed that girls' outcomes were particularly affected by early boarding.

\subsubsection{Consolidation and ethnic disparities in education}

As a group, youth who are members of officially recognized minority ethnic groups are vulnerable to consolidation due to disproportionate residence in poorer rural communities. Poverty in China is mainly concentrated in remote mountainous areas, border areas, and minority areas of central and western China \autocite[66]{Liu_Liu_Zhou_2017}. \textcite[93]{Hou_Wu_Song_2020} note that villages in central and western regions are often scattered in mountainous areas with poor access to transportation, which necessitates boarding at school\footnote{Separate from boarding associated with the left-behind children phenomenon (\chinese{留守儿童}) and school consolidation, there have been ``dislocated" boarding schools (\chinese{內地班}) for minority students from Xinjiang and Tibet in inland China for many years \autocite{Zhu_2007, Postiglione_2008, Chen_2008, Grose_2010, Chen_2013, Leibold_2019, Postiglione_Li_2021}. These schools are sometimes referred to as ``inland" or ``hinterland" schools or classes, or Xinjiang or Tibet classes.} when there are closures of village primary schools or teaching points.  A recent policy review indicated that by 2007, about 18 percent of primary students boarded, and the proportion was higher in western regions \autocite[76]{sude_policy_2020}. The review also reported that boarders in compulsory education are mainly located in large ethnic minority provinces and autonomous regions \autocite[76]{sude_policy_2020}.\footnote{In 2017, 10.66 million primary school students were boarders.  The vast majority--9.3 million--were in rural areas, and most of these in central and western China \autocite[121]{NWCCW_UNICEF_NBS_2018}.}

Sude and his colleagues also suggest that consolidation policies have paid insufficient attention to the particular educational circumstances and needs in minority communities \autocite{sude_policy_2020}. For example, consolidation may imply a shift in language of instruction for some minority youth.
\textcite{Rehamo_Harrell_2018} conducted fieldwork on bilingual education models in use in the Liangshan Yi Autonomous Prefecture in Sichuan Province in 2016, including ``first model'' schools using minority language of instruction and ``second model'' schools offering minority language classes. With official statistics, the authors trace a rapid decline in ``first model'' schools, evident in the late 1990s and accelerated in the 2000s, which the authors speculate may be linked to consolidations already underway in this area in the 1990s.\footnote{\textcite{Rehamo_Harrell_2018} also observed that schools in which ``first model'' instruction was abolished were then counted as `second model' schools, with the result
that the superficial numbers of `second model' schools did not change much since the 1990s.} This trend is also consistent with a refocusing of language policy to stress learning the national language in the twenty-first century, relative to the 1990s \autocite[16-17]{Zhang_Cai_2020}.

To the extent that consolidation implied a shift away from minority language use in schools,\footnote{Linguistic mismatch at school can occur not only because children who speak minority languages at home come into contact with Mandarin at school.  Children from smaller minority groups may come together in schools where the dominant language is that of larger minority group.  In addition, because of the ways that ethnic boundaries were drawn in China, subgroups of some minorities may speak a language other than the standard language recognized for their group, and may come into contact with a dominant group language at larger schools.  Finally, Han children who speak topolects mutually unintelligible with standard Mandarin are, in effect, not receiving mother tongue education in schools. For a discussion of linguistic diversity among Mongolian, Yi and Tibetan populations, see \textcite{roche_articulating_2019}. See \textcite{harrell_linguistics_1993} and \textcite{dwyer_texture_1998} for a discussion of linguistic diversity in the Han population.} language difficulties might negatively affect the foundations of learning for children from minority backgrounds (for example, see \textcite{wang_chinas_2015} on the experiences of Dongxiang children starting school). Language difficulties may greatly reduce the benefits of schooling \autocite[238]{lu_who_2016}. In a study in northwest China, \textcite{yang_han-minority_2015} found a minority disadvantage in school performance that was much more pronounced among language minority students than among minority students from groups that did not have a separate mother tongue.  Further, the performance of linguistic minority children benefited less from the same school inputs in mixed schools, which lacked minority language classes, compared to their Han counterparts. \textcite[10]{Rehamo_Harrell_2018} reported that about 39 percent of the teachers they surveyed in 2016 agreed or strongly agreed that inadequate knowledge of Mandarin posed an obstacle to learning for some students.  Language difficulties are likely to be particularly pronounced for students from isolated minority villages, who would have had fewer opportunities to have been exposed to the national language before starting school.\footnote{A Mandarin learning environment could be perceived in a variety of ways in minority communities. For example, \citeauthor{Hansen_1999b}'s fieldwork in the mid-1990s among Tai (Dai), Naxi, Akha (Hani) and Jinuo communities in Yunnan Province found, among other factors, that historical relations with the Han majority and the dominant minority group in the region, and group experiences of educational mobility and the lack thereof, shaped the degree to which minority youth and families found Mandarin-language education to be instrumental, beneficial, or subtractive \autocite{Hansen_1999b}.}  Separate from issues related to learning foundations, parents might be wary of sending young children to board at a school in which communication is a challenge.\footnote{Apart from effects on educational outcomes, there may also be cultural diffusion effects of school consolidation.  \textcite{liu_measuring_2019} found that gender gaps in risk-taking become more similar for Mosuo children, who come from a matriarchal culture, and Han children, for whom patriarchal families are the norm, when they are placed in school together or assigned to board together in middle school.}

From a different perspective, it is conceivable that  better-resourced schools teaching in the national language associated with consolidation might lead to stronger Mandarin skills and, ultimately, greater educational attainment. \textcite{you_language_2018} investigated impact on education and employment outcomes of a nationwide language education reform—-the Chinese Pinyin Act of 1958-1960—-using a difference-in-differences approach by interacting a birth cohort exposure dummy with the linguistic distances between local languages and standard Mandarin. The paper suggests that the policy shift led to modest short-term negative effects, but long-term positive effects, on educational attainment, and that it increased rural households’ non-agricultural employment and improved workers' capacity to migrate across provinces and language regions.  Earlier work using the Chinese Household Ethnic Survey (CHES) data employed here shows that Mandarin ability and  minority language ability are associated with quite distinct employment opportunities in young adulthood, with Mandarin ability more closely linked to economic opportunity \autocite{gustafsson_linguistic_2021}.

Finally, the benefits of access to consolidated schools might also be unequally distributed, as suggested by a case study of a 2000 elementary school system consolidation in a county in a poor, mountainous minority area in southwestern Sichuan \autocite{chan_harrell_2009}.
Through review of aggregate data and comparisons of the experiences of five elementary schools, the authors report that while consolidation increased the average quality of basic education, schools became less accessible to students
living in remote areas—-consolidation exacerbated polarization between remote, likely poorer, villages, compared to villages in the county periphery and developing areas in the county core.

 \subsection{Research questions}
In short, an emerging cross-national literature suggests that closures tend to produce larger, better-resourced schools that are farther away.  In the context of China, consolidation implies a higher likelihood of requiring boarding.
Consolidation may also imply a stronger orientation to the national language and less access to mother-tongue education, for some minority youth.  Some studies in China and elsewhere suggest that girls' enrollment might be particularly vulnerable to greater distances to school.

For minority youth, the potential benefits of better resourced schools might be tempered if consolidation implies a shift in language of instruction and language of communication, because educational engagement might be reduced, or because parents may perceive higher risk associated with children living at school when communication is a barrier.  Some international studies suggest that early grade mother-tongue education has positive impacts on educational, language, and economic outcomes, but evidence is mixed. Minority children from more demographically isolated villages might experience the greatest challenges from a more Mandarin-oriented educational environment, while one case study suggests that children from more economically marginalized villages might have lower access than others in the same county to consolidated schools when village schools close.

With this context in mind, we address three research questions: 1) How are schools serving closure villages different from schools in villages with no reported closures?  2) Compared to non-minority counterparts, are youth from officially-recognized minority groups differently affected by the experience of school closure, in terms of policy impact on both educational attainment and facility in the national language? And, 3) Are there important treatment effect heterogeneities across groups defined at the intersections of minority status, gender, and community ethnic composition and socioeconomic status? \section{Data and methods}
\subsection{Study site and sample\label{sec:data:sample}}
\label{sec:data}
This paper utilizes data from the rural household and village questionnaires from the China Household Ethnic Survey (CHES 2011), which covers households and villages from 728 villages in 81 counties of 7 provinces with substantial minority populations in China.\footnote{Please see \textcite{gustafsson_methodological_2020} for an authoritative description of the survey and discussion of some of the key variables collected.
\textcite{ches_2011} provides detailed source data in Chinese.
\textcite{Howell_2017} and \textcite{hannum_estimating_2020} provide detailed discussions on survey design and the distribution of survey villages across provinces.} In order to investigate the economic and social conditions of people in minority areas, CHES 2011 utilized subsamples of the National Bureau of Statistics' Rural Household Survey (RHS) in seven Chinese provinces and autonomous regions with substantial minority populations. Household information at the end of 2011 was collected through diaries and single-round visits in early 2012. Village closure information, along with information on the closest primary school to the village, is taken from a village head survey, which was collected in conjunction with household surveys.

For our analysis, to evaluate educational attainment and capture variations in closure effects, we focus on a sub-sample of the CHES survey with individuals between 10 to 34 years of age in 2011.\footnote{We do not consider individuals below 10 years of age in 2011 who have just started their educational trajectories. We also do not consider individuals beyond 34 years of age in 2011, who were born before market reforms.} Additionally, for individuals from villages that experienced school closure, we consider only those who were older than 5 years of age at the time of closure.\footnote{Given closure dates and the survey time-frame in 2011, 82 percent of individuals who were below 6 years of age at the year of closure are below 10 years of age in 2011. Given that we focus on individuals above 10 years of age in 2011, we do not include individuals who were below 6 years of age at the year of closure.
} Given our study selection criteria, we consider 638 villages from 80 counties in the 7 provinces.\footnote{In each of these 638 villages, there were at least two surveyed individuals that were between 10 to 34 years of age in 2011.}

\subsection{Measurement\label{sec:measurement}}
\begin{table}[htbp]
\centering
\def\sym#1{\ifmmode^{#1}\else\(^{#1}\)\fi}
\caption{Means of individual-level variables by cohort (age in 2011) and ethnic category\label{tab:stats:indi:allminohan}}
\begin{adjustbox}{max width=1\textwidth}
\begin{tabular}{m{7cm} >{\centering\arraybackslash}m{1.5cm} >{\centering\arraybackslash}m{1.5cm} >{\centering\arraybackslash}m{1.5cm} >{\centering\arraybackslash}m{1.5cm} >{\centering\arraybackslash}m{1.5cm}}
\toprule
&
\multicolumn{5}{c}{\small
Age
in
2011}
\\
\cmidrule(l{5pt}r{5pt}){2-6}
&
\multicolumn{1}{C{1.5cm}}{\textbf{\small
10-14}}
&
\multicolumn{1}{C{1.5cm}}{\textbf{\small
15-19}}
&
\multicolumn{1}{C{1.5cm}}{\textbf{\small
20-24}}
&
\multicolumn{1}{C{1.5cm}}{\textbf{\small
25-29}}
&
\multicolumn{1}{C{1.5cm}}{\textbf{\small
30-34}}
\\
\midrule
\multicolumn{6}{L{14.5cm}}{\vspace*{-5mm}\hspace*{-21mm}\textbf{{\normalsize Panel A: All individuals}}} \\&            &            &            &            &            \\
Male fraction       &        0.51&        0.53&        0.52&        0.55&        0.54\\
Currently enrolled fraction&        0.93&        0.56&        0.15&        0.02&        0.00\\
Years of education  &        5.52&        9.37&        9.91&        8.60&        7.59\\
Strong spoken Mandarin fraction&        0.20&        0.34&        0.36&        0.27&        0.22\\
Strong written Mandarin fraction&        0.16&        0.34&        0.30&        0.20&        0.16\\
\midrule
Observations        &        1879&        2392&        3153&        2425&        1788\\

\midrule
\multicolumn{6}{L{14.5cm}}{\vspace*{-5mm}\hspace*{-21mm}\textbf{{\normalsize Panel B: Minority individuals}}} \\&            &            &            &            &            \\
Male fraction       &        0.53&        0.52&        0.52&        0.55&        0.55\\
Currently enrolled fraction&        0.93&        0.53&        0.12&        0.01&        0.00\\
Years of education  &        5.41&        9.06&        9.46&        8.34&        7.34\\
Strong spoken Mandarin fraction&        0.17&        0.28&        0.29&        0.22&        0.17\\
Strong written Mandarin fraction&        0.14&        0.28&        0.23&        0.16&        0.13\\
\midrule
Observations        &        1308&        1594&        2087&        1675&        1209\\

\midrule
\multicolumn{6}{L{14.5cm}}{\vspace*{-5mm}\hspace*{-21mm}\textbf{{\normalsize Panel C: Han individuals}}} \\&            &            &            &            &            \\
Male fraction       &        0.49&        0.54&        0.51&        0.54&        0.52\\
Currently enrolled fraction&        0.93&        0.64&        0.21&        0.02&        0.01\\
Years of education  &        5.76&        9.98&       10.78&        9.18&        8.13\\
Strong spoken Mandarin fraction&        0.27&        0.47&        0.49&        0.37&        0.34\\
Strong written Mandarin fraction&        0.21&        0.45&        0.44&        0.29&        0.22\\
\midrule
Observations        &         571&         798&        1066&         750&         579\\
\bottomrule
\addlinespace[-1.5em]
\multicolumn{6}{L{16.5cm}}{
\footnotesize
\justify
Individual-level information is based on responses from the household respondent. See Section \ref{sec:data:sample} for details on the study site and sample. See Section \ref{sec:measurement} for discussions on the educational outcomes and demographic variables shown in Table \ref{tab:stats:indi:allminohan}.
}\\
\end{tabular}
\end{adjustbox}
\end{table}
 \begin{table}[htbp]
\centering
\caption{Means (s.d.) of village and closest primary school variables (2011)\label{tab:stats:meansd:vil}}
\begin{adjustbox}{max width=1\textwidth}
\begin{tabular}{m{8cm} >{\centering\arraybackslash}m{1.625cm} >{\centering\arraybackslash}m{1.625cm} >{\centering\arraybackslash}m{1.625cm} >{\centering\arraybackslash}m{1.625cm}}
\toprule
&
\multicolumn{1}{c}{\small
}
&
\multicolumn{3}{c}{\small
Han
fraction
inclusion
threshold}
\\
\cmidrule(l{5pt}r{5pt}){2-2}
\cmidrule(l{5pt}r{5pt}){3-5}
&
\multicolumn{1}{C{1.625cm}}{All
villages}
&
\multicolumn{1}{C{1.625cm}}{{\small
Han
$<90$\%}}
&
\multicolumn{1}{C{1.625cm}}{{\small
Han
$<70$\%}}
&
\multicolumn{1}{C{1.625cm}}{{\small
Han
$<50$\%}}
\\
\midrule
\multicolumn{5}{p{14.5cm}}{\vspace*{-1mm}\hspace*{0mm}\textbf{Panel A: Closure, school distance and quality}} \\
\addlinespace
Village reports school closure&        0.27&        0.25&        0.24&        0.24\\
                    &\vspace*{-2mm}{\footnotesize (0.45) }&\vspace*{-2mm}{\footnotesize (0.44) }&\vspace*{-2mm}{\footnotesize (0.43) }&\vspace*{-2mm}{\footnotesize (0.43) }\\
Distance measure from village head survey&        2.71&        2.54&        2.49&        2.54\\
                    &\vspace*{-2mm}{\footnotesize (5.18) }&\vspace*{-2mm}{\footnotesize (5.28) }&\vspace*{-2mm}{\footnotesize (5.24) }&\vspace*{-2mm}{\footnotesize (5.49) }\\
Primary school facilities index&        5.37&        5.22&        5.19&        5.21\\
                    &\vspace*{-2mm}{\footnotesize (1.99) }&\vspace*{-2mm}{\footnotesize (1.93) }&\vspace*{-2mm}{\footnotesize (1.93) }&\vspace*{-2mm}{\footnotesize (1.90) }\\
Has no unsafe buildings&        0.82&        0.81&        0.82&        0.83\\
                    &\vspace*{-2mm}{\footnotesize (0.39) }&\vspace*{-2mm}{\footnotesize (0.39) }&\vspace*{-2mm}{\footnotesize (0.39) }&\vspace*{-2mm}{\footnotesize (0.38) }\\
\midrule
\multicolumn{5}{p{14.5cm}}{\vspace*{-1mm}\hspace*{0mm}\textbf{Panel B: School language offerings and facilities}} \\
\addlinespace
Has teachers from Han ethnic group&        0.81&        0.75&        0.72&        0.68\\
                    &\vspace*{-2mm}{\footnotesize (0.40) }&\vspace*{-2mm}{\footnotesize (0.44) }&\vspace*{-2mm}{\footnotesize (0.45) }&\vspace*{-2mm}{\footnotesize (0.47) }\\
Has teachers from minority ethnic groups&        0.82&        0.93&        0.96&        0.97\\
                    &\vspace*{-2mm}{\footnotesize (0.39) }&\vspace*{-2mm}{\footnotesize (0.26) }&\vspace*{-2mm}{\footnotesize (0.20) }&\vspace*{-2mm}{\footnotesize (0.18) }\\
Has minority language classes&        0.22&        0.25&        0.27&        0.29\\
                    &\vspace*{-2mm}{\footnotesize (0.41) }&\vspace*{-2mm}{\footnotesize (0.44) }&\vspace*{-2mm}{\footnotesize (0.44) }&\vspace*{-2mm}{\footnotesize (0.45) }\\
Has minority language of instruction&        0.27&        0.32&        0.35&        0.39\\
                    &\vspace*{-2mm}{\footnotesize (0.44) }&\vspace*{-2mm}{\footnotesize (0.47) }&\vspace*{-2mm}{\footnotesize (0.48) }&\vspace*{-2mm}{\footnotesize (0.49) }\\
\midrule
\multicolumn{5}{p{14.5cm}}{\vspace*{-1mm}\hspace*{0mm}\textbf{Panel C: School boarding facilities}} \\
\addlinespace
Requires living away from home&        0.31&        0.32&        0.32&        0.32\\
                    &\vspace*{-2mm}{\footnotesize (0.46) }&\vspace*{-2mm}{\footnotesize (0.47) }&\vspace*{-2mm}{\footnotesize (0.47) }&\vspace*{-2mm}{\footnotesize (0.47) }\\
Has dormitory       &        0.39&        0.40&        0.40&        0.40\\
                    &\vspace*{-2mm}{\footnotesize (0.49) }&\vspace*{-2mm}{\footnotesize (0.49) }&\vspace*{-2mm}{\footnotesize (0.49) }&\vspace*{-2mm}{\footnotesize (0.49) }\\
\midrule
\multicolumn{5}{p{14.5cm}}{\vspace*{-1mm}\hspace*{0mm}\textbf{Panel D: Other village attributes}} \\
\addlinespace
Number of households&       446.0&       420.0&       417.7&       421.7\\
                    &\vspace*{-2mm}{\footnotesize (305.9) }&\vspace*{-2mm}{\footnotesize (287.0) }&\vspace*{-2mm}{\footnotesize (294.7) }&\vspace*{-2mm}{\footnotesize (302.7) }\\
Fraction Han in village&        0.40&        0.20&        0.13&       0.076\\
                    &\vspace*{-2mm}{\footnotesize (0.42) }&\vspace*{-2mm}{\footnotesize (0.29) }&\vspace*{-2mm}{\footnotesize (0.21) }&\vspace*{-2mm}{\footnotesize (0.14) }\\
Fraction female in village&        0.48&        0.48&        0.48&        0.48\\
                    &\vspace*{-2mm}{\footnotesize (0.054) }&\vspace*{-2mm}{\footnotesize (0.053) }&\vspace*{-2mm}{\footnotesize (0.053) }&\vspace*{-2mm}{\footnotesize (0.053) }\\
\midrule
Observations        &         638&         477&         426&         382\\
\bottomrule
\addlinespace[-0.75em]
\multicolumn{5}{p{15.75cm}}{
\vspace*{-3mm} \footnotesize\justify
Village level information is based on responses from the village head. Primary school information is based on the closest full primary school to the village in 2011.
}\\
\end{tabular}
\end{adjustbox}
\end{table}

In this section, we first discuss individual-level educational attainment, language facility and demographic variables. We then discuss variables reported by the village head on village ethnic composition, school closure, and language offerings and facilities of the closest primary school to the village in 2011.

\subsubsection{Outcome variables}

\paragraph{Educational attainment} We measure educational attainment based on years of schooling completed for each individual at the time of the survey in 2011. Table \ref{tab:stats:indi:allminohan} shows that, in our sample, minority and Han individuals who are between 20 to 24 years of age in 2011 have on average 9.5 and 10.8 years of schooling, respectively. Individuals who are 20 to 24 years of age in 2011 have higher educational attainment compared to younger individuals who are still proceeding through school as well as to individuals from older cohorts who were born in the late 1970s and early 1980s (between 25 to 34 years of age in 2011).

A significant fraction of the younger sample is currently enrolled. The percentages enrolled are around 93, 56 and 15 percent for individuals at 10 to 14, 15 to 19 and 20 to 24 years of age in 2011, respectively. Minority enrollment is generally lower than enrollment for Han individuals. For example, in 2011, for individuals between 20 to 24 years of age, 12 percent of minority individuals and 21 percent of Han individuals were enrolled in school.

\paragraph{Reported Mandarin language facility} Mother-tongue languages in China include numerous Chinese topolects (\chinese{方言}), some of which are not mutually intelligible, as well as numerous minority languages that span a broad range of language families. Topolects share a largely common written form with the national language, while the diverse range of minority languages in China do not.

To index facility with the national language, for each individual in the household, \textit{spoken Mandarin ability} and \textit{written Mandarin ability} are reported. These variables were originally coded into four categories: strong, good, simple ability, or none.\footnote{The household survey respondent rated the spoken Mandarin communication ability and written Mandarin reading and writing ability of each household member. The questions had four categorical responses: have strong ability, have good ability, have simple ability, and no ability.
} This approach is similar to that used in the US Census, in which a report (or proxy report) is given about speaking ability (\textcite{Bureau_2019} explains the approach; \textcite{kominski_how_2014} describes past validations of the approach).\footnote{\textcite{kominski_how_2014} reported that verification exercises using the 1980 census and an English Language Proficiency Study showed that people who were reported to speak English "very well", but not others, had passing levels similar to the control group of English speakers. Census alternatives to "very well" are "well", "not well" and "not at all". This report also suggested analyses of a 1986 test census showed that language speaking capacity measures showed a general consistency between the items used in the census and a series of other language related items, including background, current use contexts, and other operational measures presumed to be less subjective than the "how well" questions. For discussion, see \textcite{Angrist_Chin_Godoy_2008}. Other work with the CHES rural data, using a combined measure of reported writing and speaking capacity, has shown that better reported Mandarin facility is associated with greater Internet use and better economic outcomes for rural minority young adults \autocite{gustafsson_linguistic_2021}.} We generated a contrast of strong reported language facility, coded as one, with all other capabilities coded as zero.\footnote{For our main analysis on Mandarin ability, we exclude non-responses which account for approximately 10 percent of the sample. In Appendix Section \ref{sec:app:esti:mandwithna}, we re-estimate Mandarin ability regressions by including non-respondents jointly with individuals reported to have good, simple, or no Mandarin ability. In Appendix Table \ref{tab:stats:indi:mandwritegroups}, we show that, conditional on age in 2011, non-respondents of the Mandarin ability question have lower enrollment and lower years of grades completed than strong Mandarin ability respondents.
}

Table \ref{tab:stats:indi:allminohan} shows that Han individuals in our sample report higher fractions with strong Mandarin language skills than minority individuals. For reading and writing ability, the fraction of Han individuals reporting strong ability is between 50 to 100 percent higher than minority individuals. In particular, individuals between 20 to 24 years of age in 2011 have the largest gap, with 23 and 44 percent of minority and Han individuals reporting having strong written Mandarin ability. Across cohorts, the strong Mandarin ability fraction doubles between individuals who were in early to later teen years in 2011, and then halves for older cohorts between 30 to 34 years of age in 2011.

\subsubsection{Child demographics and socioeconomic status}

\paragraph{Minority status} Minority status is coded 1 if a household member is a member of an officially-designated minority ethnic group, or 0 if the individual is reported as a member of the Han majority. In our analytic sample of individuals from these predominantly minority area villages, 7873 individuals belong to minority ethnic groups and 3764 individuals are members of the Han majority.\footnote{In Appendix Section \ref{sec:minobreakdown}, we provide a province-by-province breakdown of the main ethnic groups in the sample. Our sample includes eight of the ten largest minority groups in China. Manchus and Yi are the 3rd and 6th largest minority groups in China, but reside in areas outside of our sampled regions.}$^{,}$\footnote{For our main analysis, we include all minority individuals in the regression. In Appendix Section \ref{sec:app:esti:minoownlang}, we test the robustness of our results to the exclusion of minority individuals reported not to have their own language. Those excluded individuals predominantly come from the Hui minority group, who mostly reside in the Ningxia Hui Autonomous Region.}

\paragraph{Gender} The survey collected binary responses regarding sex of individuals in the household. A household member is coded 1 if reported as male, otherwise 0. Our analytic sample includes 6161 male individuals and 5476 female individuals.

\paragraph{Age cohorts} Age at the end of 2011 is reported and is used to define single-year-of-age cohorts. Models incorporate province-specific cohort fixed effects, as part of the identification strategy defined below. We focus on outcomes for individuals who are between 10 and 34 years of age in 2011. Table \ref{tab:stats:indi:allminohan} and Appendix Tables \ref{tab:stats:indi:ageatclosure}, \ref{tab:stats:indi:fourgrps}, and \ref{tab:stats:indi:mandwritegroups} present age structure, educational attainment and enrollment, and spoken and written Mandarin ability statistics for cohort groups in 2011.

\subsubsection{Closure variables}

\paragraph{School closure} School closure information is reported in the village questionnaire. Table \ref{tab:stats:meansd:vil} includes the closure variable in Panel A. Village heads were asked if the village had a primary school, and asked about the year of school closure if the village school had been closed. Among the 638 villages in the analytic sample, there are 175 closure villages that did not have village schools in 2011 and experienced school closure between 1999 and 2010. The analytic sample includes 2698 individuals from these villages with school closure.

\paragraph{Age at closure} There is heterogeneity in the timing of school closure. Out of the 175 closure villages, 20 experienced school closure between 1999 and 2002, 49 between 2003 and 2006, and 106 between 2007 and 2010.\footnote{Among the 638 villages in our analytic sample, school closure took place in all seven provinces and in all the year ranges listed. The intensity of school closure varied across provinces. Between 1999 and 2010, 46 and 53 percent of the surveyed villages from Hunan Province in south-central China and the Inner Mongolia Autonomous Region in north China reported village school closures, respectively. 24, 21, and 20 percent of surveyed villages from Ningxia Hui Autonomous Region in northwestern China, the Xinjiang Uygur Autonomous Region in northwestern China, and Guizhou Province in southwestern China experienced closures, respectively. 17 and 19 percent of surveyed villages from the Guangxi Zhuang Autonomous Region in south-central China and Qinghai Province in northwest China had school closure, respectively.} Given variations in the village-specific year of closure, there are two different dimensions of time: the year when a child was born and the year of school closure in that child’s village. We use the difference between these two variables to calculate \textit{age at closure}.
We consider individuals who were 6 to 9, 10 to 13, 14 to 21 and 22 to 29 years of age in the year of school closure. Appendix Table \ref{tab:stats:indi:ageatclosure} summarizes educational attainment and Mandarin ability by cohorts (age in 2011) and ages at years of closure. For example, among individuals from closure villages who were 10 to 13 years of age at the village-specific years of closure, 129 individuals were between ages 10 to 14 in 2011, 222 individuals were between ages 15 to 19 in 2011, and 115 individuals were over 20 years of age in 2011.

\subsubsection{Village and province context}

\paragraph{Village ethnic composition} The village questionnaire provides information on the proportion of Han individuals in the village population. Column 1 of Table \ref{tab:stats:meansd:vil} presents village and closest primary school statistics for all villages. Columns 2, 3, and 4 focus on villages with less than 90, 70, and 50 percent Han population, respectively. There are 382 villages with less than 50 percent Han individuals, 95 villages with between 50 and 90 percent Han individuals, and 161 villages with greater than 90 percent Han individuals.

\paragraph{Village school characteristics} The village questionnaire contains information on distance to school as well as nine yes-no questions about basic facilities and infrastructure for the closest primary school to the village that is attended by village children. Facility questions include whether the primary school has no unsafe buildings, has heating, tap water, kitchen, shower, sufficient desks, a library, personal computers, and internet access. We generate a \textit{facilities index} based on summing these binary variables. Panel A of Table \ref{tab:stats:meansd:vil} shows that distance and the facilities index are similar across Han fraction inclusion thresholds. Distance to the closest primary school is on average 2.71 km and 2.54 km for all villages and predominantly minority villages, respectively. Across the columns, around 82 percent of primary schools report having no unsafe buildings and the average school facilities index varies between 5.21 to 5.37 (out of 9 points maximum).

In Panel B and C of Table \ref{tab:stats:meansd:vil}, we include measures of teacher demographics (\textit{has any Han teachers}, \textit{has any minority teachers}) and boarding status (\textit{requires living away from home}, \textit{school has dormitories}). Additionally, we include measures of \textit{minority language of instruction} and \textit{minority language classes} as indicators of a minority-centered curriculum. The share of closest primary schools reported to have a Han teacher drops from 81 percent in all villages to 68 percent in predominantly minority villages. The fraction of closest primary schools having a teacher of minority ethnic background increases from 82 percent in all villages to 97 percent in predominantly minority villages. For all villages, the fraction with minority language classes and the fraction that employ a minority language as the language of instruction are 22 and 27 percent, respectively. These fractions rise to 29 and 39 percent for predominantly minority villages. Across Han fraction inclusion thresholds, 40 percent of schools have dormitories and 32 percent of schools require living away from home.

 \subsection{Methods}
\label{sec:identify}

We undertake a parallel set of analyses for two education-related outcomes: educational attainment and facility with the national language.  For individuals of the same province and age cohorts (age in 2011), we first compare the difference in educational attainment (number of grades completed by 2011) or language facility between those from closure villages who were exposed to closure to individuals from non-closure villages. This first difference is attributable to the consolidation policy as well as existing differences in educational attainment across villages that would have occurred in the absence of the policy. We make a second comparison of the difference in educational attainment between closure village individuals not impacted by the policy and individuals of the same age cohorts from non-closure villages. This second difference should only reflect existing differences in educational attainment across villages. The impact of the policy is, in effect, the difference in the first and the second differences.

\label{rr:roneidentifyone}
In closure villages, we classify individuals into groups who are exposed to the policy upon entry into primary school (6 to 9 years old at the time of primary school closure) or who experienced closure in the second-half of primary school (10 to 13 years old at the time of primary school closure). Exposed individuals are distinguished from older individuals of the same village who were above elementary school age (age 13) at the time of closure. Individuals from villages without closure do not have age at closure.

\label{rr:aethreeintro}
The primary underlying assumption of our strategy is that, in the absence of the closure policy, rural educational attainment or language facility follows a common province-specific age cohort trend. Cohort-invariant differences across villages, which might be correlated with school closure, are captured by village-specific fixed effects.\footnote{Controlling for provincial fixed effects, \textcite{hannum_estimating_2020} find no statistically significant differences between closure and non-closure villages in terms of village per-capita income, local wage, fraction of workers in the labor force, the fraction of migrant workers, village per-capita spending overall and on education, and the implementation rate of health and insurance policies between closure and non-closure villages. The key statistically significant difference is that closure villages have on average 11 percent less households than non-closure villages.} To test for pre-existing cohort trends, we separate older individuals from closure villages into those who were 14 to 21 and those who were 22 to 29 at the year of closure. In the absence of cohort trends, the difference in attainment or language facility between each of these age-at-closure groups and individuals of the same age cohorts (age in 2011) from non-closure villages should be similar.\footnote{In Appendix Section \ref{sec:app:stats:attain}, we compare educational attainment and strong Mandarin ability between individuals from non-closure villages and individuals from closure villages who were not exposed to closure. We find that differences in raw averages disappear once we control for province fixed effects, age fixed effects, and gender.} Additional concerns might arise if individuals or families move as a result of the closure policy. In the rural China context, however, mandatory household registration limited mobility of entire households.
Additionally, the CHES survey, conducted around the New Year holidays, captures the educational status of all household members including migrant workers.

\label{sec:identifyageeffect}
Equation \eqref{eq:startsOnly} presents the estimation equation. Our regression approach assumes that in the scenario without policy intervention, educational outcome, \(E\)---which might be number of grades completed or reported language facility---of a child \(i\) from village \(v\) in province \(p\) and whose cohort (age in 2011) is \(a\) could be decomposed into five parts: a village fixed effect \(\beta_{v}\), a province-specific cohort fixed effect $\gamma_{pa}$, an ethnic-minority-specific cohort fixed effect $\delta_{a}$, and idiosyncratic terms including one part that can be explained by observed characteristic \(X_{i}\) and another unobserved error term \(\epsilon_{i}\).  $M_i$ indicates whether an individual is Han ($M_i = 0$) or minority ($M_i = 1$). Under school closure, the policy's effect is assumed to be additive and captured by \(\theta^{H}_{z}\) for Han individuals and \(\theta^{M}_{z}\) for minorities, both of which are specific to group \(z\) defined by age-at-closure \(t_{i}\):

\label{rr:rtwoqoneageeffect}
\label{rr:roneqtwoageeffects}
\begin{singlespace}\vspace*{-\baselineskip}
\begin{eqnarray}
E_{pvia} & = & \alpha + \beta_{v} + \gamma_{pa} + \delta_{a}\cdot M_i \nonumber \\
&  & + \sum_{z=1}^Z \left(\theta^{H}_{z}\cdot \left(1-M_i\right) + \theta^{M}_{z}\cdot M_i \right)\cdot \mathbf{1}\left\{ l_{z}\leq t_{i}\leq u_{z}\right\} \cdot C_{v}\label{eq:startsOnly}\\
&  & +X_i\cdot\gamma\nonumber +\epsilon_{i}\nonumber
\end{eqnarray}
\end{singlespace}\noindent\ignorespaces
In Equation \eqref{eq:startsOnly}, \(\alpha\) is a constant, and \(C_{v}\) is a binary variable indicating if individual \(i\) is from a village \(v\) with school consolidation (i.e. treatment village). We group children in villages with school closure into \(Z\) groups based on their age at closure \(t_{i}\), with lower and upper bounds for each group \(l_{z}\) and \(u_{z}\). Therefore \(\theta^{H}_{z}\) and \(\theta^{M}_{z}\) capture the average effects across cohorts for age-at-closure group \(z\) at the time of survey measurement. For individuals from villages without closure, \(C_{v}=0\). Including individuals from villages without closure isolates the policy effects from the province-specific cohort patterns captured by \(\rho_{pa}\).\footnote{The specification assumes that $\theta_{z}$ is not specific to 2011 age $a$ and $\theta_{z}$ is not specific to the calendar year of closure. Additionally, we use common age-at-closure cut-offs, $l_{z}$ and $u_{z}$ for all demographic groups.}

Our second research question addressed whether there was a differential impact of closure by ethnic minority status. We address this question by comparing the estimated policy effects \(\theta^{H}_{z}\) for Han individuals and \(\theta^{M}_{z}\) for minorities. Educational outcome cohort trajectories might differ for Han and minority youth for reasons that are unrelated to the consolidation policy, which the inclusion of $\delta_a$ helps to account for.

Our third research question asked whether there are important treatment effect heterogeneities across groups defined at the intersections of minority status, gender, and community ethnic composition and economic status. To allow for the possibility that more ethnically segregated minority areas were differently affected by closures, we estimate
policy effects separately for predominantly Han villages and predominantly minority villages. To estimate ethnic gaps in closure impact conditioned on gender and community economic status, we estimate Equation \eqref{eq:startsOnly} separately for sex groups and obtain group-specific coefficients--\(\theta^{H,g}_{z}\) and \(\theta^{M, g}_{z}\)--for age-at-closure and sex groups, with $g\in\left\{\text{male}, \text{female}\right\}$. The village fixed effects and provincial-cohort trends are allowed to be gender-specific. We show estimates for all villages as well as more minority-centric villages. Additionally, we also split the sample into villages with below and above median average per-capita income levels, and estimate sex group-specific regressions for relatively poorer and richer villages separately. This approach, in effect,  constitutes a treatment effect heterogeneity analysis in which the intersections of minority status, gender, and village ethnic composition and economic status might delineate substantially distinct vulnerabilities to policy effects.

\section{Results}
\subsection{Schools serving closure- and non-closure villages}
\begin{table}[htbp]
\centering
\def\sym#1{\ifmmode^{#1}\else\(^{#1}\)\fi}
\caption{Means of closest primary school variables (2011), closure and non-closure villages\label{tab:stats:byclosure:pval}}
\begin{adjustbox}{max width=1\textwidth}
\begin{tabular}{m{8cm} >{\centering\arraybackslash}m{1.625cm} >{\centering\arraybackslash}m{1.625cm} >{\centering\arraybackslash}m{1.625cm} >{\centering\arraybackslash}m{1.625cm}}
\toprule
&
\multicolumn{2}{C{3.25cm}}{\small
\textbf{Group
averages}}
&
\multicolumn{2}{C{3.25cm}}{\small
\textbf{P-values}}
\\
\cmidrule(l{5pt}r{5pt}){2-3}
\cmidrule(l{5pt}r{5pt}){4-5}
&
\multicolumn{1}{C{1.625cm}}{{\small
Non-closure}}
&
\multicolumn{1}{C{1.625cm}}{{\small
Closure}}
&
\multicolumn{1}{C{1.625cm}}{{\small
Closure
vs
non-closure\dag}}
&
\multicolumn{1}{C{1.625cm}}{{\small
Year
of
closure
trend\ddag}}
\\
\midrule
\multicolumn{5}{L{14.5cm}}{\vspace*{-5mm}\hspace*{-17mm}\textbf{\normalsize Panel A: All villages}}\\&            &            &            &            \\
Distance (km) to primary school&        1.80&        5.41&        0.00&        0.15\\
Primary school facilities index (0 to 9)&        5.07&        6.28&        0.00&        0.65\\
Has no unsafe buildings&        0.79&        0.90&        0.00&        0.40\\
\cmidrule(l{5pt}r{10pt}){1-1}Has teachers from Han ethnic group&        0.75&        0.96&        0.00&        0.02\\
Has teachers from minority ethnic groups&        0.83&        0.79&        0.71&        0.91\\
Has minority language classes&        0.23&        0.16&        0.56&        0.07\\
Has minority language of instruction&        0.31&        0.15&        0.00&        0.67\\
\cmidrule(l{5pt}r{10pt}){1-1}Requires living away from home&        0.25&        0.50&        0.00&        0.22\\
Has dormitory       &        0.31&        0.62&        0.00&        0.38\\
\midrule
\multicolumn{5}{L{14.5cm}}{\vspace*{-5mm}\hspace*{-17mm}\textbf{\normalsize Panel B: Villages where Han fraction is less than 50 percent}}\\&            &            &            &            \\
Distance (km) to primary school&        1.73&        5.24&        0.00&        0.15\\
Primary school facilities index (0 to 9)&        4.91&        6.20&        0.00&        0.65\\
Has no unsafe buildings&        0.80&        0.91&        0.00&        0.40\\
\cmidrule(l{5pt}r{10pt}){1-1}Has teachers from Han ethnic group&        0.61&        0.93&        0.00&        0.02\\
Has teachers from minority ethnic groups&        0.97&        0.94&        0.71&        0.91\\
Has minority language classes&        0.31&        0.22&        0.56&        0.07\\
Has minority language of instruction&        0.43&        0.22&        0.00&        0.67\\
\cmidrule(l{5pt}r{10pt}){1-1}Requires living away from home&        0.26&        0.49&        0.00&        0.22\\
Has dormitory       &        0.33&        0.63&        0.00&        0.38\\
\bottomrule
\addlinespace[-1.5em]
\multicolumn{5}{L{16.1cm}}{
\footnotesize
\justify
Village level information is based on responses from the village head. Primary school information is based on the closest full primary school to the village in 2011. {\large\dag} Column 3 tests whether the mean differences between non-closure and closure villages are statistically significant after controlling for provincial fixed effects. {\large\ddag} Column 4 tests whether the means within closure villages are statistically different across years of closure, also controlling for provincial fixed effects.
}\\
\end{tabular}
\end{adjustbox}
\end{table}

Are the pathways ``better'' in consolidated schools? We can not directly estimate the causal effects of closure on facilities and curricular offering because we do not observe how facilities and curricular offerings changed with closure. We can observe differences in school characteristics in villages that have and have not experienced closure in 2011. Table \ref{tab:stats:byclosure:pval} shows descriptive mean comparisons of key closure and language related variables reported by the village head for the closest primary school to the village between villages that experienced closure and villages that did not experience closure. We present means in columns 1 and 2. In column 3, we test whether the mean differences are statistically significant after controlling for provincial fixed effects. In column 4, we test whether the means within closure villages are statistically different across years of closure, also controlling for provincial fixed effects. In separate panels, we present statistics for all villages and villages where the Han fraction is less than 50 percent.

Table \ref{tab:stats:byclosure:pval} suggests, as expected, that villages reporting closures also report schools that are more distant, but have better resources than non-closure villages. In Panel A, across all villages types, the average distance to school reported for closure villages was 5.78 kilometers, while the average distance to school reported for non-closure villages was 1.81 kilometers. This difference is highly statistically significant. Closure villages in the sample report better facilities. Out of a maximum index value of 9, closure villages have an average aggregate school facilities index value of 6.07 compared to 5.09 for non-closure villages. One of the component measures of the index is whether school has no unsafe buildings. For this measure, we find that 19 percent of non-closure villages report that the closest primary schools have unsafe buildings, compared to 9 percent for closure villages.\footnote{The means for seven of the nine individual facility measures that generate the facilities index are statistically different between closure and non-closure villages. The exceptions are having sufficient desks and having access to tap water. For all nine measures, the share of closure villages with the facility is larger than the share of non-closure villages with it.} We do not find significant time-trends among closure villages for distance and facility measures. Results in Panel B for villages where the Han fraction is less than 50 percent are similar to statistics for all villages.

Table \ref{tab:stats:byclosure:pval} also presents means for teacher demographics, linguistic and cultural focus, and boarding facilities. In Panels A and B, there is a pattern of significantly greater likelihood of schools having a Han teacher, significantly less likelihood of schools having minority language of instruction, and significantly greater likelihood that schools requiring boarding for schools serving closure villages. Closure village schools are also somewhat less likely to have minority language classes, but this difference, after controlling for provincial fixed effects, is not significant. Additionally, in the sample, closure villages and non-closure villages report schools with similarly high levels of having teachers from minority ethnic groups. For all variables, the gaps between closure and non-closure villages are similar across village level of segregation, though more ethnically isolated minority villages in the sample have lower rates of Han teachers and higher rates of minority teachers and language offerings.

Given these descriptive findings, consolidated schools tend to be better than other village schools in terms of resources, but due to teacher demographics, curricular offerings, and greater distance and the need to board, consolidated schools may have been viewed by some minority parents as both less oriented toward cultures and languages of minority students and more risky for young children without fluency in the national language. However, it is also possible that the prospect of stronger national language acquisition might be particularly valued by language minority parents, due to its instrumentality for economic mobility.
 \subsection{Heterogeneity of closure impact by minority status and village ethnic composition}
\begin{table}[htbp]
\centering
\caption{Regressions of educational attainment with school closure, Han-minority interaction\label{tab:interhan:attain:nsfto}}
\begin{adjustbox}{max width=1\textwidth}
\begin{tabular}{m{8.5cm} >{\centering\arraybackslash}m{1.5cm} >{\centering\arraybackslash}m{1.5cm} >{\centering\arraybackslash}m{1.5cm} >{\centering\arraybackslash}m{1.5cm} >{\centering\arraybackslash}m{1.5cm} >{\centering\arraybackslash}m{1.5cm}}
\toprule
&
\multicolumn{6}{c}{Outcome:
grades
completed
by
year
2011}
\\
\cmidrule(l{5pt}r{5pt}){2-7}
&
\multicolumn{1}{c}{\small
}
&
\multicolumn{5}{c}{
Village
Han
fraction
inclusion
threshold}
\\
\cmidrule(l{5pt}r{5pt}){2-2}
\cmidrule(l{5pt}r{5pt}){3-7}
&
\multicolumn{1}{C{1.5cm}}{All
villages}
&
\multicolumn{1}{C{1.5cm}}{Han
$<90$\%}
&
\multicolumn{1}{C{1.5cm}}{Han
$<70$\%}
&
\multicolumn{1}{C{1.5cm}}{Han
$<50$\%}
&
\multicolumn{1}{C{1.5cm}}{Han
$<10$\%}
&
\multicolumn{1}{C{1.5cm}}{Han
$\ge
90$\%}
\\
\midrule
\multicolumn{7}{L{19cm}}{\vspace*{-3.5mm}\hspace*{-8mm}\textit{Baseline group: Child was 14-21 years old at village primary school closure year}} \\            \multicolumn{7}{L{19cm}}{\vspace*{-5mm}\hspace*{-8mm}{Child is minority interaction:}} \\&                     &                     &                     &                     &                     &                     \\
\vspace*{0mm}\hspace*{5mm} Closure $\times$ age at closure 6-9&       -0.50\sym{**} &       -0.54\sym{**} &       -0.45\sym{*}  &       -0.44\sym{*}  &       -0.46         &                     \\
                    &\vspace*{-2mm}{\footnotesize (0.23) }         &\vspace*{-2mm}{\footnotesize (0.24) }         &\vspace*{-2mm}{\footnotesize (0.25) }         &\vspace*{-2mm}{\footnotesize (0.26) }         &\vspace*{-2mm}{\footnotesize (0.31) }         &                     \\
\vspace*{0mm}\hspace*{5mm} Closure $\times$ age at closure 10-13&       -0.60\sym{***}&       -0.66\sym{***}&       -0.56\sym{***}&       -0.58\sym{***}&       -0.56\sym{**} &                     \\
                    &\vspace*{-2mm}{\footnotesize (0.19) }         &\vspace*{-2mm}{\footnotesize (0.19) }         &\vspace*{-2mm}{\footnotesize (0.19) }         &\vspace*{-2mm}{\footnotesize (0.20) }         &\vspace*{-2mm}{\footnotesize (0.24) }         &                     \\
\vspace*{0mm}\hspace*{5mm} Closure $\times$ age at closure 22-29&       0.076         &       0.025         &        0.12         &        0.10         &       0.012         &                     \\
                    &\vspace*{-2mm}{\footnotesize (0.19) }         &\vspace*{-2mm}{\footnotesize (0.19) }         &\vspace*{-2mm}{\footnotesize (0.20) }         &\vspace*{-2mm}{\footnotesize (0.21) }         &\vspace*{-2mm}{\footnotesize (0.28) }         &                     \\
\multicolumn{7}{L{19cm}}{\vspace*{-5mm}\hspace*{-8mm}{Child is Han interaction:}} \\&                     &                     &                     &                     &                     &                     \\
\vspace*{0mm}\hspace*{5mm} Closure $\times$ age at closure 6-9&      0.0058         &       -0.45         &      -0.065         &       -0.60         &                     &        0.40         \\
                    &\vspace*{-2mm}{\footnotesize (0.26) }         &\vspace*{-2mm}{\footnotesize (0.39) }         &\vspace*{-2mm}{\footnotesize (0.41) }         &\vspace*{-2mm}{\footnotesize (0.55) }         &                     &\vspace*{-2mm}{\footnotesize (0.32) }         \\
\vspace*{0mm}\hspace*{5mm} Closure $\times$ age at closure 10-13&       -0.28         &       -0.25         &       0.010         &       -0.52         &                     &       -0.20         \\
                    &\vspace*{-2mm}{\footnotesize (0.20) }         &\vspace*{-2mm}{\footnotesize (0.32) }         &\vspace*{-2mm}{\footnotesize (0.41) }         &\vspace*{-2mm}{\footnotesize (0.38) }         &                     &\vspace*{-2mm}{\footnotesize (0.26) }         \\
\vspace*{0mm}\hspace*{5mm} Closure $\times$ age at closure 22-29&       -0.19         &      -0.087         &     0.00036         &      -0.077         &                     &       -0.15         \\
                    &\vspace*{-2mm}{\footnotesize (0.25) }         &\vspace*{-2mm}{\footnotesize (0.31) }         &\vspace*{-2mm}{\footnotesize (0.38) }         &\vspace*{-2mm}{\footnotesize (0.40) }         &                     &\vspace*{-2mm}{\footnotesize (0.39) }         \\
\midrule
\hspace*{3mm}Observations        &       11637         &        9197         &        8371         &        7653         &        5686         &        2451         \\
\bottomrule
\addlinespace[-1.5em]
\multicolumn{7}{L{19.7cm}}{
\footnotesize\justify
Statistical significance:* 0.10 ** 0.05 *** 0.01. Robust standard error clustered at village level. Each column represents a separate regression. All regressions include village fixed effects, province-specific age FEs, Han-specific age FEs, and control for household size. Columns 2 to 5 sequentially drop villages with more than 90, 70, 50, and 10 percent Han population. Samples for columns 5 and 6 include only minority and Han individuals, respectively.
}\\
\end{tabular}
\end{adjustbox}
\end{table}
 \begin{table}[htbp]
\centering
\caption{Regressions of Mandarin ability with school closure, Han-minority status interaction\label{tab:interhan:mandwritetalkm2c:nsfto}}
\begin{adjustbox}{max width=1\textwidth}
\begin{tabular}{m{8.5cm} >{\centering\arraybackslash}m{1.5cm} >{\centering\arraybackslash}m{1.5cm} >{\centering\arraybackslash}m{1.5cm} >{\centering\arraybackslash}m{1.5cm} >{\centering\arraybackslash}m{1.5cm} >{\centering\arraybackslash}m{1.5cm}}
\toprule
&
\multicolumn{6}{c}{Outcome:
strong
Mandarin
ability
=
1,
no/simple/good
=
0}
\\
\cmidrule(l{5pt}r{5pt}){2-7}
&
\multicolumn{1}{c}{\small
}
&
\multicolumn{5}{c}{
Village
Han
fraction
inclusion
threshold}
\\
\cmidrule(l{5pt}r{5pt}){2-2}
\cmidrule(l{5pt}r{5pt}){3-7}
&
\multicolumn{1}{C{1.5cm}}{All
villages}
&
\multicolumn{1}{C{1.5cm}}{Han
$<90$\%}
&
\multicolumn{1}{C{1.5cm}}{Han
$<70$\%}
&
\multicolumn{1}{C{1.5cm}}{Han
$<50$\%}
&
\multicolumn{1}{C{1.5cm}}{Han
$<10$\%}
&
\multicolumn{1}{C{1.5cm}}{Han
$\ge
90$\%}
\\
\midrule
\addlinespace[+1em]
\multicolumn{7}{L{19cm}}{\vspace*{1mm}\hspace*{-8mm}\textbf{Panel A: Written Mandarin ability}} \\
\multicolumn{7}{L{19cm}}{\vspace*{-3.5mm}\hspace*{-8mm}\textit{Baseline group: Child was 14-21 years old at village primary school closure year}} \\
\multicolumn{7}{L{19cm}}{\vspace*{-5mm}\hspace*{-8mm}{Child is minority interaction:}} \\
&                     &                     &                     &                     &                     &                     \\
\vspace*{0mm}\hspace*{5mm} Closure $\times$ age at closure 6-9&       -0.14\sym{***}&       -0.15\sym{***}&       -0.16\sym{***}&       -0.18\sym{***}&       -0.23\sym{***}&                     \\
                    &\vspace*{-2mm}{\footnotesize (0.044) }         &\vspace*{-2mm}{\footnotesize (0.045) }         &\vspace*{-2mm}{\footnotesize (0.047) }         &\vspace*{-2mm}{\footnotesize (0.048) }         &\vspace*{-2mm}{\footnotesize (0.057) }         &                     \\
\vspace*{0mm}\hspace*{5mm} Closure $\times$ age at closure 10-13&      -0.054\sym{*}  &      -0.067\sym{**} &      -0.067\sym{**} &      -0.073\sym{**} &      -0.073\sym{*}  &                     \\
                    &\vspace*{-2mm}{\footnotesize (0.031) }         &\vspace*{-2mm}{\footnotesize (0.031) }         &\vspace*{-2mm}{\footnotesize (0.031) }         &\vspace*{-2mm}{\footnotesize (0.033) }         &\vspace*{-2mm}{\footnotesize (0.040) }         &                     \\
\vspace*{0mm}\hspace*{5mm} Closure $\times$ age at closure 22-29&      -0.016         &      -0.020         &      -0.013         &      -0.018         &      -0.061         &                     \\
                    &\vspace*{-2mm}{\footnotesize (0.032) }         &\vspace*{-2mm}{\footnotesize (0.032) }         &\vspace*{-2mm}{\footnotesize (0.034) }         &\vspace*{-2mm}{\footnotesize (0.036) }         &\vspace*{-2mm}{\footnotesize (0.047) }         &                     \\
\multicolumn{7}{L{19cm}}{\vspace*{-5mm}\hspace*{-8mm}{Child is Han interaction:}} \\&                     &                     &                     &                     &                     &                     \\
\vspace*{0mm}\hspace*{5mm} Closure $\times$ age at closure 6-9&       0.076         &      -0.041         &      -0.035         &      -0.018         &                     &        0.18\sym{**} \\
                    &\vspace*{-2mm}{\footnotesize (0.055) }         &\vspace*{-2mm}{\footnotesize (0.074) }         &\vspace*{-2mm}{\footnotesize (0.098) }         &\vspace*{-2mm}{\footnotesize (0.14) }         &                     &\vspace*{-2mm}{\footnotesize (0.075) }         \\
\vspace*{0mm}\hspace*{5mm} Closure $\times$ age at closure 10-13&      -0.034         &      -0.026         &      -0.027         &      -0.059         &                     &      -0.043         \\
                    &\vspace*{-2mm}{\footnotesize (0.043) }         &\vspace*{-2mm}{\footnotesize (0.061) }         &\vspace*{-2mm}{\footnotesize (0.093) }         &\vspace*{-2mm}{\footnotesize (0.12) }         &                     &\vspace*{-2mm}{\footnotesize (0.058) }         \\
\vspace*{0mm}\hspace*{5mm} Closure $\times$ age at closure 22-29&      -0.022         &       0.020         &      -0.052         &      -0.075         &                     &      -0.034         \\
                    &\vspace*{-2mm}{\footnotesize (0.038) }         &\vspace*{-2mm}{\footnotesize (0.045) }         &\vspace*{-2mm}{\footnotesize (0.048) }         &\vspace*{-2mm}{\footnotesize (0.063) }         &                     &\vspace*{-2mm}{\footnotesize (0.056) }         \\
\midrule
\hspace*{3mm}Observations        &       10495         &        8126         &        7328         &        6656         &        4745         &        2383         \\

\midrule
\addlinespace[+1em]
\multicolumn{7}{L{19cm}}{\vspace*{1mm}\hspace*{-8mm}\textbf{Panel B: Spoken Mandarin ability}} \\
\multicolumn{7}{L{19cm}}{\vspace*{-3.5mm}\hspace*{-8mm}\textit{Baseline group: Child was 14-21 years old at village primary school closure year}} \\
\multicolumn{7}{L{19cm}}{\vspace*{-5mm}\hspace*{-8mm}{Child is minority interaction:}} \\
&                     &                     &                     &                     &                     &                     \\
\vspace*{0mm}\hspace*{5mm} Closure $\times$ age at closure 6-9&      -0.035         &      -0.033         &      -0.030         &      -0.043         &      -0.078         &                     \\
                    &\vspace*{-2mm}{\footnotesize (0.049) }         &\vspace*{-2mm}{\footnotesize (0.051) }         &\vspace*{-2mm}{\footnotesize (0.053) }         &\vspace*{-2mm}{\footnotesize (0.055) }         &\vspace*{-2mm}{\footnotesize (0.076) }         &                     \\
\vspace*{0mm}\hspace*{5mm} Closure $\times$ age at closure 10-13&      -0.014         &      -0.024         &      -0.020         &      -0.023         &      -0.027         &                     \\
                    &\vspace*{-2mm}{\footnotesize (0.033) }         &\vspace*{-2mm}{\footnotesize (0.034) }         &\vspace*{-2mm}{\footnotesize (0.036) }         &\vspace*{-2mm}{\footnotesize (0.037) }         &\vspace*{-2mm}{\footnotesize (0.052) }         &                     \\
\vspace*{0mm}\hspace*{5mm} Closure $\times$ age at closure 22-29&      -0.015         &      -0.011         &     -0.0072         &      -0.010         &      -0.061         &                     \\
                    &\vspace*{-2mm}{\footnotesize (0.029) }         &\vspace*{-2mm}{\footnotesize (0.030) }         &\vspace*{-2mm}{\footnotesize (0.031) }         &\vspace*{-2mm}{\footnotesize (0.032) }         &\vspace*{-2mm}{\footnotesize (0.042) }         &                     \\
\multicolumn{7}{L{19cm}}{\vspace*{-5mm}\hspace*{-8mm}{Child is Han interaction:}} \\&                     &                     &                     &                     &                     &                     \\
\vspace*{0mm}\hspace*{5mm} Closure $\times$ age at closure 6-9&       0.053         &      -0.042         &       0.028         &      -0.050         &                     &        0.11         \\
                    &\vspace*{-2mm}{\footnotesize (0.052) }         &\vspace*{-2mm}{\footnotesize (0.075) }         &\vspace*{-2mm}{\footnotesize (0.072) }         &\vspace*{-2mm}{\footnotesize (0.098) }         &                     &\vspace*{-2mm}{\footnotesize (0.070) }         \\
\vspace*{0mm}\hspace*{5mm} Closure $\times$ age at closure 10-13&     -0.0028         &      -0.040         &       0.052         &      -0.038         &                     &      0.0096         \\
                    &\vspace*{-2mm}{\footnotesize (0.035) }         &\vspace*{-2mm}{\footnotesize (0.057) }         &\vspace*{-2mm}{\footnotesize (0.069) }         &\vspace*{-2mm}{\footnotesize (0.10) }         &                     &\vspace*{-2mm}{\footnotesize (0.048) }         \\
\vspace*{0mm}\hspace*{5mm} Closure $\times$ age at closure 22-29&     -0.0021         &       0.040         &      0.0023         &      0.0100         &                     &      -0.013         \\
                    &\vspace*{-2mm}{\footnotesize (0.040) }         &\vspace*{-2mm}{\footnotesize (0.056) }         &\vspace*{-2mm}{\footnotesize (0.068) }         &\vspace*{-2mm}{\footnotesize (0.089) }         &                     &\vspace*{-2mm}{\footnotesize (0.053) }         \\
\midrule
\hspace*{3mm}Observations        &       10488         &        8126         &        7332         &        6662         &        4745         &        2373         \\
\bottomrule
\addlinespace[-1.5em]
\multicolumn{7}{L{19.7cm}}{
\footnotesize\justify
Statistical significance:* 0.10 ** 0.05 *** 0.01. Robust standard error clustered at village level. Each column in each Panel represents a separate linear probability regression. All regressions include village fixed effects, province-specific age FEs, Han-specific age FEs, and control for household size. Columns 2 to 5 sequentially drop villages with more than 90, 70, 50, and 10 percent Han population. Samples for columns 5 and 6 include only minority and Han individuals, respectively. For these regressions, we consider individuals who responded to the written or spoken Mandarin ability questions.
}\\
\end{tabular}
\end{adjustbox}
\end{table}

In Tables \ref{tab:interhan:attain:nsfto} and \ref{tab:interhan:mandwritetalkm2c:nsfto}, we present the impacts of closure on educational attainment and Mandarin language ability for children who were 6 to 9 and 10 to 12 years of age at the time of closure, along with a ``placebo test'' of a closure effect for individuals who were 22 to 29 at the time of closure. In both tables, we interact age at the time of closure with whether a child is Han or minority. Moving left to right from columns one to four in both tables, we present estimates that focus on successively more minority-centric villages. For example, column 1 contains all villages, and column 4 does not include majority-Han villages. In column 5, we include only minority children from nearly all minority villages by considering villages where the fraction of Han individuals is less than 10 percent. In column 6, we include only Han children from nearly all Han villages by considering villages where the fraction of Han individuals is greater than 90 percent.

Because models include ethnicity- and province-specific single-year-of-age cohort controls as well as village fixed effects, the age-at-closure reference category, 14 to 21 year olds, accommodates any base differences in closure villages compared to non-closure villages, but for \textit{unaffected cohorts}--those for whom primary school closure should not have had any impact. The age-at-closure coefficients for ages 6 to 9 and 10 to 13, in turn, offer the decrement in years of schooling \textit{in affected cohorts} in affected villages, compared to the years of schooling predicted by village and cohort fixed-effects. It is this quantity, estimated for different groups, that we will refer to in the discussions below as ``the decrement" and treat as the policy effect.\footnote{As shown in Appendix Table \ref{tab:stats:indi:ageatclosure}, in 2011, due to differential school closure dates, individuals who were 6 to 9 at year of closure predominantly are between 10 to 19 years of age in 2011. Individuals who were 10 to 13 at year of closure predominantly are between 10 to 24 years of age in 2011. The policy effect is the average effects on years of schooling from both individuals who have possibly completed schooling as well as individuals who are still progressing through school.} Finally, a term for age at closure of 22 to 29 is also estimated and presented.

Table \ref{tab:interhan:attain:nsfto} analyzes closure effects on educational attainment, with cohort-specific closure effects allowed to vary with minority status of the child. Results show that, by 2011, minority youth ages 10 to 13 at time of closure in a closure village experienced around a 0.6-year decrement in years of schooling. For those 6 to 9 years old at closure, the decrement is around a half-year. Results are similar in more and less minority-centric villages, with a weakening of statistical significance for estimates from columns based on smaller numbers of villages. In contrast to the situation of minority youth, the same set of estimates for Han youth shows no statistically significant decrements in attainment associated with experiencing school closure. For both minority and Han individuals, we do not find a closure effect for the placebo group of individuals that were 22 to 29 at the year of closure.

We ran a parallel set of linear probability regressions for closure effects on reported ability in written and spoken Mandarin in Table \ref{tab:interhan:mandwritetalkm2c:nsfto}.\footnote{Results in Table \ref{tab:interhan:mandwritetalkm2c:nsfto} exclude non-respondents. We find very similar results in Appendix Table \ref{tab:interhan:mandwritetalkm2c:nsfto:withna} where we include non-respondents in the same group as individuals who answered no/simple/good to the Mandarin ability question.} Results for written Mandarin ability in the top panel show decrements in the probability of reporting strong written ability for those affected by closures at the beginning of their primary school careers. The estimated decrement for minority individuals is larger in the more isolated samples: 14 percentage points in the most integrated villages compared to 23 percentage points in the most isolated minority villages. There is a smaller decrement in reported written ability for those exposed to closure at ages 10 to 13, which suggests about a 5 to 7 percentage points reduction in the likelihood of reporting strong Mandarin written ability. Significant decrements are not observed for Han children across village groups. On the contrary, in the most Han-centric villages in column 6, a 18 percentage points \textit{increase} in the probability of reporting strong written Mandarin ability is reported for Han children who started in consolidated schools. For reported spoken Mandarin communication ability, shown in the bottom panel of Table \ref{tab:interhan:mandwritetalkm2c:nsfto}, we find no significant effects. For this reason, subsequent analyses focus exclusively on attainment and written Mandarin.

 \subsection{Intersectional perspectives: Adding gender}
\begin{table}[htbp]
\centering
\caption{Regressions of educational attainment with school closure, Han-minority status interaction by gender\label{tab:interhan:bygender:attain:nsfto}}
\begin{adjustbox}{max width=1\textwidth}
\begin{tabular}{m{8.5cm} >{\centering\arraybackslash}m{1.5cm} >{\centering\arraybackslash}m{1.5cm} >{\centering\arraybackslash}m{1.5cm} >{\centering\arraybackslash}m{1.5cm} >{\centering\arraybackslash}m{1.5cm} >{\centering\arraybackslash}m{1.5cm}}
\toprule
&
\multicolumn{6}{c}{Outcome:
grades
completed
by
year
2011}
\\
\cmidrule(l{5pt}r{5pt}){2-7}
&
\multicolumn{3}{c}{\textbf{
Female}}
&
\multicolumn{3}{c}{\textbf{
Male}}
\\
\cmidrule(l{5pt}r{5pt}){2-4}
\cmidrule(l{5pt}r{5pt}){5-7}
&
\multicolumn{1}{C{1.5cm}}{All
villages}
&
\multicolumn{1}{C{1.5cm}}{Han
$<90$\%
villages}
&
\multicolumn{1}{C{1.5cm}}{Han
$<50$\%
villages}
&
\multicolumn{1}{C{1.5cm}}{All
villages}
&
\multicolumn{1}{C{1.5cm}}{Han
$<90$\%
villages}
&
\multicolumn{1}{C{1.5cm}}{Han
$<50$\%
villages}
\\
\midrule
\multicolumn{7}{L{19cm}}{\vspace*{-3.5mm}\hspace*{-8mm}\textit{Baseline group: Child was 14-21 years old at village primary school closure year}} \\            \multicolumn{7}{L{19cm}}{\vspace*{-5mm}\hspace*{-8mm}{Child is minority interaction:}} \\&                     &                     &                     &                     &                     &                     \\
\vspace*{0mm}\hspace*{5mm} Closure $\times$ age at closure 6-9&       -0.58         &       -0.55         &       -0.26         &       -0.19         &       -0.31         &       -0.25         \\
                    &\vspace*{-2mm}{\footnotesize (0.36) }         &\vspace*{-2mm}{\footnotesize (0.35) }         &\vspace*{-2mm}{\footnotesize (0.39) }         &\vspace*{-2mm}{\footnotesize (0.26) }         &\vspace*{-2mm}{\footnotesize (0.26) }         &\vspace*{-2mm}{\footnotesize (0.30) }         \\
\vspace*{0mm}\hspace*{5mm} Closure $\times$ age at closure 10-13&       -0.77\sym{**} &       -0.75\sym{**} &       -0.63         &       -0.49\sym{**} &       -0.62\sym{***}&       -0.72\sym{***}\\
                    &\vspace*{-2mm}{\footnotesize (0.32) }         &\vspace*{-2mm}{\footnotesize (0.33) }         &\vspace*{-2mm}{\footnotesize (0.38) }         &\vspace*{-2mm}{\footnotesize (0.24) }         &\vspace*{-2mm}{\footnotesize (0.24) }         &\vspace*{-2mm}{\footnotesize (0.26) }         \\
\vspace*{0mm}\hspace*{5mm} Closure $\times$ age at closure 22-29&        0.23         &        0.23         &        0.13         &       0.083         &       0.031         &        0.18         \\
                    &\vspace*{-2mm}{\footnotesize (0.30) }         &\vspace*{-2mm}{\footnotesize (0.31) }         &\vspace*{-2mm}{\footnotesize (0.33) }         &\vspace*{-2mm}{\footnotesize (0.28) }         &\vspace*{-2mm}{\footnotesize (0.29) }         &\vspace*{-2mm}{\footnotesize (0.30) }         \\
\multicolumn{7}{L{19cm}}{\vspace*{-5mm}\hspace*{-8mm}{Child is Han interaction:}} \\&                     &                     &                     &                     &                     &                     \\
\vspace*{0mm}\hspace*{5mm} Closure $\times$ age at closure 6-9&       -0.35         &       -0.46         &                     &        0.36         &       -0.55         &                     \\
                    &\vspace*{-2mm}{\footnotesize (0.34) }         &\vspace*{-2mm}{\footnotesize (0.51) }         &                     &\vspace*{-2mm}{\footnotesize (0.41) }         &\vspace*{-2mm}{\footnotesize (0.53) }         &                     \\
\vspace*{0mm}\hspace*{5mm} Closure $\times$ age at closure 10-13&       -0.45         &       -0.36         &                     &       0.018         &       -0.11         &                     \\
                    &\vspace*{-2mm}{\footnotesize (0.35) }         &\vspace*{-2mm}{\footnotesize (0.55) }         &                     &\vspace*{-2mm}{\footnotesize (0.27) }         &\vspace*{-2mm}{\footnotesize (0.34) }         &                     \\
\vspace*{0mm}\hspace*{5mm} Closure $\times$ age at closure 22-29&       -0.23         &       0.019         &                     &       0.013         &       -0.32         &                     \\
                    &\vspace*{-2mm}{\footnotesize (0.33) }         &\vspace*{-2mm}{\footnotesize (0.42) }         &                     &\vspace*{-2mm}{\footnotesize (0.37) }         &\vspace*{-2mm}{\footnotesize (0.46) }         &                     \\
\midrule
\hspace*{3mm}Observations        &        5476         &        4328         &        3328         &        6161         &        4869         &        3830         \\
\bottomrule
\addlinespace[-1.5em]
\multicolumn{7}{L{19.7cm}}{
\footnotesize\justify
Statistical significance:* 0.10 ** 0.05 *** 0.01. Robust standard error clustered at village level. Each column represents a separate regression. All regressions include village fixed effects, province-specific age FEs, Han-specific age FEs, and control for household size. Columns 1, 2, and 3 regressions include only females. Columns 4, 5, and 6 regressions include only males. Columns 2 to 3 (and 5 to 6) sequentially drop villages with more than 90 and 50 percent Han population. Samples for columns 3 and 6 regressions include only minority individuals.
}\\
\end{tabular}
\end{adjustbox}
\end{table}
 \begin{table}[htbp]
\centering
\caption{Regressions of written Mandarin ability with school closure, Han-minority status interaction by gender\label{tab:interhan:bygender:mandwritem2c:nsfto}}
\begin{adjustbox}{max width=1\textwidth}
\begin{tabular}{m{8.5cm} >{\centering\arraybackslash}m{1.5cm} >{\centering\arraybackslash}m{1.5cm} >{\centering\arraybackslash}m{1.5cm} >{\centering\arraybackslash}m{1.5cm} >{\centering\arraybackslash}m{1.5cm} >{\centering\arraybackslash}m{1.5cm}}
\toprule
&
\multicolumn{6}{c}{{\small Outcome:
strong
written
Mandarin
ability
=
1,
no/simple/good
=
0}}
\\
\cmidrule(l{5pt}r{5pt}){2-7}
&
\multicolumn{3}{c}{\textbf{
Female}}
&
\multicolumn{3}{c}{\textbf{
Male}}
\\
\cmidrule(l{5pt}r{5pt}){2-4}
\cmidrule(l{5pt}r{5pt}){5-7}
&
\multicolumn{1}{C{1.5cm}}{All
villages}
&
\multicolumn{1}{C{1.5cm}}{Han
$<90$\%
villages}
&
\multicolumn{1}{C{1.5cm}}{Han
$<50$\%
villages}
&
\multicolumn{1}{C{1.5cm}}{All
villages}
&
\multicolumn{1}{C{1.5cm}}{Han
$<90$\%
villages}
&
\multicolumn{1}{C{1.5cm}}{Han
$<50$\%
villages}
\\
\midrule
\multicolumn{7}{L{19cm}}{\vspace*{-3.5mm}\hspace*{-8mm}\textit{Baseline group: Child was 14-21 years old at village primary school closure year}} \\            \multicolumn{7}{L{19cm}}{\vspace*{-5mm}\hspace*{-8mm}{Child is minority interaction:}} \\&                     &                     &                     &                     &                     &                     \\
\vspace*{0mm}\hspace*{5mm} Closure $\times$ age at closure 6-9&       -0.13\sym{**} &       -0.12\sym{**} &       -0.14\sym{**} &       -0.12\sym{**} &       -0.14\sym{**} &       -0.19\sym{***}\\
                    &\vspace*{-2mm}{\footnotesize (0.056) }         &\vspace*{-2mm}{\footnotesize (0.057) }         &\vspace*{-2mm}{\footnotesize (0.061) }         &\vspace*{-2mm}{\footnotesize (0.053) }         &\vspace*{-2mm}{\footnotesize (0.055) }         &\vspace*{-2mm}{\footnotesize (0.056) }         \\
\vspace*{0mm}\hspace*{5mm} Closure $\times$ age at closure 10-13&       -0.11\sym{**} &       -0.13\sym{***}&       -0.12\sym{**} &      -0.013         &      -0.028         &      -0.046         \\
                    &\vspace*{-2mm}{\footnotesize (0.047) }         &\vspace*{-2mm}{\footnotesize (0.047) }         &\vspace*{-2mm}{\footnotesize (0.050) }         &\vspace*{-2mm}{\footnotesize (0.042) }         &\vspace*{-2mm}{\footnotesize (0.044) }         &\vspace*{-2mm}{\footnotesize (0.048) }         \\
\vspace*{0mm}\hspace*{5mm} Closure $\times$ age at closure 22-29&     -0.0050         &     -0.0032         &      -0.033         &      -0.016         &      -0.026         &      -0.022         \\
                    &\vspace*{-2mm}{\footnotesize (0.038) }         &\vspace*{-2mm}{\footnotesize (0.036) }         &\vspace*{-2mm}{\footnotesize (0.040) }         &\vspace*{-2mm}{\footnotesize (0.047) }         &\vspace*{-2mm}{\footnotesize (0.049) }         &\vspace*{-2mm}{\footnotesize (0.051) }         \\
\multicolumn{7}{L{19cm}}{\vspace*{-5mm}\hspace*{-8mm}{Child is Han interaction:}} \\&                     &                     &                     &                     &                     &                     \\
\vspace*{0mm}\hspace*{5mm} Closure $\times$ age at closure 6-9&       0.028         &      -0.049         &                     &        0.12         &      -0.019         &                     \\
                    &\vspace*{-2mm}{\footnotesize (0.071) }         &\vspace*{-2mm}{\footnotesize (0.11) }         &                     &\vspace*{-2mm}{\footnotesize (0.079) }         &\vspace*{-2mm}{\footnotesize (0.094) }         &                     \\
\vspace*{0mm}\hspace*{5mm} Closure $\times$ age at closure 10-13&      -0.074         &      -0.087         &                     &      0.0020         &       0.020         &                     \\
                    &\vspace*{-2mm}{\footnotesize (0.059) }         &\vspace*{-2mm}{\footnotesize (0.082) }         &                     &\vspace*{-2mm}{\footnotesize (0.056) }         &\vspace*{-2mm}{\footnotesize (0.073) }         &                     \\
\vspace*{0mm}\hspace*{5mm} Closure $\times$ age at closure 22-29&     0.00091         &     -0.0054         &                     &      -0.044         &       0.010         &                     \\
                    &\vspace*{-2mm}{\footnotesize (0.050) }         &\vspace*{-2mm}{\footnotesize (0.067) }         &                     &\vspace*{-2mm}{\footnotesize (0.056) }         &\vspace*{-2mm}{\footnotesize (0.066) }         &                     \\
\midrule
\hspace*{3mm}Observations        &        4917         &        3800         &        2865         &        5578         &        4326         &        3333         \\
\bottomrule
\addlinespace[-1.5em]
\multicolumn{7}{L{19.7cm}}{
\footnotesize\justify
Statistical significance:* 0.10 ** 0.05 *** 0.01. Robust standard error clustered at village level. Each column in each Panel represents a separate linear probability regression. All regressions include village fixed effects, province-specific age FEs, Han-specific age FEs, and control for household size. Columns 1, 2, and 3 regressions include only females. Columns 4, 5, and 6 regressions include only males. Columns 2 to 3 (and 5 to 6) sequentially drop villages with more than 90 and 50 percent Han population. Samples for columns 3 and 6 regressions include only minority individuals.
}\\
\end{tabular}
\end{adjustbox}
\end{table}

Building on the results from the last section, we next consider whether the significant negative closure impacts on educational attainment and writing facility observed among minority youth are borne equally by boys and girls. In Tables \ref{tab:interhan:bygender:attain:nsfto} and \ref{tab:interhan:bygender:mandwritem2c:nsfto}, we present the impacts of closure on educational attainment and mandarin language writing abilities for females and males, Han and ethnic minority children.\footnote{In both tables, to account for differences in gender that are unrelated to closure, we estimate the model for females and males separately, which allows for different village fixed effects for females and males as well as different ethnicity and province-specific age fixed effects.} Conditional on gender, following Tables \ref{tab:interhan:attain:nsfto} and \ref{tab:interhan:mandwritetalkm2c:nsfto}, we interact age at the time of closure with whether a child is Han or minority. We present results for females in the first three columns and results for males in the latter three columns. Our sample includes all villages in columns 1 and 4, villages with less than 90 percent Han individuals in columns 2 and 5, and less than 50 percent Han individuals in columns 3 and 6. Given the small number of female and male Han children that remain for the restricted sample of villages with less than 50 percent Han children, we include only minority children in the estimation samples for columns 3 and 6.

By 2011, Table \ref{tab:interhan:bygender:attain:nsfto} shows that closure decrements for minority girls who were ages 10 to 13 at time of closure are about three quarters of a year, except for a non-significant decrement among the most isolated villages at about 0.63 years. For boys who were ages 10 to 13 at time of closure, the pattern shows about a half-year decrement among youth in all villages; this number becomes 0.72 in the most ethnically segregated minority villages. For girls and boys who were 6 to 9 at time of closure, similar to the results in Table \ref{tab:interhan:attain:nsfto}, closure effects are negative as well, but small in magnitude compared to individuals who were older at time of closure. No significant findings are present among Han youth, although the direction of effects tends to be relatively more negative for Han girls. Thus, in the sample of all villages, minority girls are most vulnerable, but in the most ethnically segregated villages, minority boys become more vulnerable. For minority youth across genders, we find weakly positive but not significant closure effects for the placebo group of individuals.

Considering the analysis of reported writing facility in Table \ref{tab:interhan:bygender:mandwritem2c:nsfto}, from linear probability regressions, we observe a consistent decrement for minority girls in the 6 to 9 and 10 to 13 age groups of 11 to 13 percentage points, with no pattern across villages.\footnote{Table \ref{tab:interhan:bygender:mandwritem2c:nsfto} results exclude non-respondents. Appendix Table \ref{tab:interhan:bygender:mandwritem2c:nsfto:withna} shows similar estimates when non-respondents are included in the same group as individuals who answered no/simple/good to the Mandarin ability question.} However, for boys, the pattern is different. Only the youngest boys at closure are affected, and the decrement is much larger in the most ethnically isolated villages (about 19 percentage points, compared to 12 percentage points for all villages). We do not find significant results for Han youth, although the magnitude and direction of the effects for Han girls are more negative than Han boys.

 \subsection{Intersectional perspectives: Adding village income}
\begin{table}[htbp]
\centering
\caption{Regressions of educational attainment and written Mandarin ability with school closure, Han-minority status interaction by gender and village-income\label{tab:interhan:bygender:byvilrich:attain:nsfto}}
\begin{adjustbox}{max width=1\textwidth}
\begin{tabular}{m{7.5cm} >{\centering\arraybackslash}m{2.5cm} >{\centering\arraybackslash}m{2.5cm} >{\centering\arraybackslash}m{2.5cm} >{\centering\arraybackslash}m{2.5cm}}
\toprule
&
\multicolumn{4}{c}{Outcomes
in
2011}
\\
\cmidrule(l{5pt}r{5pt}){2-5}
&
\multicolumn{2}{c}{\small
Grades
completed}
&
\multicolumn{2}{c}{\small
Strong
written
Mandarin}
\\
\cmidrule(l{5pt}r{5pt}){2-3}
\cmidrule(l{5pt}r{5pt}){4-5}
&
\multicolumn{1}{C{2.5cm}}{Female}
&
\multicolumn{1}{C{2.5cm}}{Male}
&
\multicolumn{1}{C{2.5cm}}{Female}
&
\multicolumn{1}{C{2.5cm}}{Male}
\\
\midrule
\addlinespace[+1em]
\multicolumn{5}{L{17.5cm}}{\vspace*{1mm}\hspace*{-14mm}\textbf{Panel A: Villages with below 4000 Yuan per-capita income in 2011}} \\
\multicolumn{5}{L{17.5cm}}{\vspace*{-3.5mm}\hspace*{-14mm}\textit{Baseline group: Child was 14-21 years old at village primary school closure year}} \\
\multicolumn{5}{L{17.5cm}}{\vspace*{-5mm}\hspace*{-14mm}Child is minority interaction:} \\
&                     &                     &                     &                     \\
\vspace*{0mm}\hspace*{5mm} Closure $\times$ age at closure 6-9&       -1.08\sym{**} &       -0.25         &       -0.12\sym{*}  &       -0.18\sym{***}\\
                    &\vspace*{-2mm}{\footnotesize (0.45) }         &\vspace*{-2mm}{\footnotesize (0.34) }         &\vspace*{-2mm}{\footnotesize (0.070) }         &\vspace*{-2mm}{\footnotesize (0.066) }         \\
\vspace*{0mm}\hspace*{5mm} Closure $\times$ age at closure 10-13&       -1.11\sym{**} &       -0.69\sym{**} &       -0.12         &     -0.0045         \\
                    &\vspace*{-2mm}{\footnotesize (0.51) }         &\vspace*{-2mm}{\footnotesize (0.32) }         &\vspace*{-2mm}{\footnotesize (0.074) }         &\vspace*{-2mm}{\footnotesize (0.057) }         \\
\vspace*{0mm}\hspace*{5mm} Closure $\times$ age at closure 22-29&       0.060         &       0.020         &      -0.038         &      -0.045         \\
                    &\vspace*{-2mm}{\footnotesize (0.41) }         &\vspace*{-2mm}{\footnotesize (0.26) }         &\vspace*{-2mm}{\footnotesize (0.047) }         &\vspace*{-2mm}{\footnotesize (0.056) }         \\
\multicolumn{5}{L{17.5cm}}{\vspace*{-5mm}\hspace*{-14mm}{Child is Han interaction:}} \\&                     &                     &                     &                     \\
\vspace*{0mm}\hspace*{5mm} Closure $\times$ age at closure 6-9&       -0.31         &        0.35         &       0.064         &        0.17         \\
                    &\vspace*{-2mm}{\footnotesize (0.65) }         &\vspace*{-2mm}{\footnotesize (0.71) }         &\vspace*{-2mm}{\footnotesize (0.097) }         &\vspace*{-2mm}{\footnotesize (0.12) }         \\
\vspace*{0mm}\hspace*{5mm} Closure $\times$ age at closure 10-13&       -0.23         &       0.075         &        0.11         &      -0.079         \\
                    &\vspace*{-2mm}{\footnotesize (0.49) }         &\vspace*{-2mm}{\footnotesize (0.35) }         &\vspace*{-2mm}{\footnotesize (0.089) }         &\vspace*{-2mm}{\footnotesize (0.075) }         \\
\vspace*{0mm}\hspace*{5mm} Closure $\times$ age at closure 22-29&       0.014         &       -0.61         &      -0.022         &      -0.087         \\
                    &\vspace*{-2mm}{\footnotesize (0.44) }         &\vspace*{-2mm}{\footnotesize (0.41) }         &\vspace*{-2mm}{\footnotesize (0.079) }         &\vspace*{-2mm}{\footnotesize (0.088) }         \\
\midrule
\hspace*{3mm}Observations        &        2867         &        3253         &        2593         &        2980         \\

\midrule
\addlinespace[+1em]
\multicolumn{5}{L{17.5cm}}{\vspace*{1mm}\hspace*{-14mm}\textbf{Panel B: Villages with above 4000 Yuan per-capita income in 2011}} \\
\multicolumn{5}{L{17.5cm}}{\vspace*{-3.5mm}\hspace*{-14mm}\textit{Baseline group: Child was 14-21 years old at village primary school closure year}} \\
\multicolumn{5}{L{17.5cm}}{\vspace*{-5mm}\hspace*{-14mm}Child is minority interaction:} \\
&                     &                     &                     &                     \\
\vspace*{0mm}\hspace*{5mm} Closure $\times$ age at closure 6-9&       0.011         &       0.018         &       -0.14         &      -0.032         \\
                    &\vspace*{-2mm}{\footnotesize (0.59) }         &\vspace*{-2mm}{\footnotesize (0.41) }         &\vspace*{-2mm}{\footnotesize (0.086) }         &\vspace*{-2mm}{\footnotesize (0.090) }         \\
\vspace*{0mm}\hspace*{5mm} Closure $\times$ age at closure 10-13&       -0.33         &       -0.46         &      -0.075         &      -0.040         \\
                    &\vspace*{-2mm}{\footnotesize (0.43) }         &\vspace*{-2mm}{\footnotesize (0.39) }         &\vspace*{-2mm}{\footnotesize (0.052) }         &\vspace*{-2mm}{\footnotesize (0.070) }         \\
\vspace*{0mm}\hspace*{5mm} Closure $\times$ age at closure 22-29&        0.49         &       0.061         &       0.056         &       0.035         \\
                    &\vspace*{-2mm}{\footnotesize (0.42) }         &\vspace*{-2mm}{\footnotesize (0.64) }         &\vspace*{-2mm}{\footnotesize (0.066) }         &\vspace*{-2mm}{\footnotesize (0.080) }         \\
\multicolumn{5}{L{17.5cm}}{\vspace*{-6mm}\hspace*{-14mm}{Child is Han interaction:}} \\&                     &                     &                     &                     \\
\vspace*{0mm}\hspace*{5mm} Closure $\times$ age at closure 6-9&       -0.44         &        0.44         &      -0.022         &        0.16         \\
                    &\vspace*{-2mm}{\footnotesize (0.44) }         &\vspace*{-2mm}{\footnotesize (0.57) }         &\vspace*{-2mm}{\footnotesize (0.099) }         &\vspace*{-2mm}{\footnotesize (0.10) }         \\
\vspace*{0mm}\hspace*{5mm} Closure $\times$ age at closure 10-13&       -0.56         &      -0.069         &       -0.17\sym{**} &       0.064         \\
                    &\vspace*{-2mm}{\footnotesize (0.48) }         &\vspace*{-2mm}{\footnotesize (0.37) }         &\vspace*{-2mm}{\footnotesize (0.076) }         &\vspace*{-2mm}{\footnotesize (0.075) }         \\
\vspace*{0mm}\hspace*{5mm} Closure $\times$ age at closure 22-29&       -0.14         &        0.64         &       0.040         &     0.00075         \\
                    &\vspace*{-2mm}{\footnotesize (0.49) }         &\vspace*{-2mm}{\footnotesize (0.62) }         &\vspace*{-2mm}{\footnotesize (0.066) }         &\vspace*{-2mm}{\footnotesize (0.068) }         \\
\midrule
\hspace*{3mm}Observations        &        2609         &        2908         &        2324         &        2598         \\
\bottomrule
\addlinespace[-1.5em]
\multicolumn{5}{L{18.85cm}}{
\footnotesize\justify
Statistical significance:* 0.10 ** 0.05 *** 0.01. Robust standard error clustered at village level. Each column in each Panel represents a separate linear regression. All regressions include village fixed effects, province-specific age FEs, Han-specific age FEs, and control for household size. Columns 1 and 3 include only females. Columns 2 and 4 include only males. Columns 1 to 2 have educational attainment (grades completed by year 2011) as the outcome variable. Columns 3 and 4 have strong written Mandarin ability (strong = 1, no/simple/good = 0) in 2011 as the outcome variable.
}\\
\end{tabular}
\end{adjustbox}
\end{table}

In this section, we study heterogeneous effects of closure for richer and poorer villages. Building on our prior results, within village groups, we continue to allow for heterogeneous effects for minority and Han children and by gender. In Table \ref{tab:interhan:bygender:byvilrich:attain:nsfto}, in the top panel, we present results for villages with lower than 4000 Yuan per-capita income in 2011. In the bottom panel, results from separate regressions are shown for villages with higher than 4000 Yuan per-capita income in 2011.\footnote{Running separate regressions for relatively richer and poorer villages allow for different ethnicity- and provincial-specific age fixed effects for relatively richer and poorer villages.} Among the villages we study, 4000 Yuan was approximately the median village per-capita income level in 2011. We present estimates for educational attainment in columns 1 and 2 and estimates for written Mandarin ability in columns 3 and 4. To ensure sufficient sample sizes, we limit the analysis here to all villages.

In columns 1 and 2 of Table \ref{tab:interhan:bygender:byvilrich:attain:nsfto}, it is very clear that decrements to educational attainment with school closure emerge most strongly among minority youth in the relatively poorer villages. Negative attainment effects are more consistent for minority girls, shown in the top panel of column 1. Minority girls who experienced closure at ages 6 to 9 as well as ages 10 to 13 experienced around a year decrement. Among minority youth, boys who started school with closure experienced no decrement, but those who were likely in school when it closed (ages 10 to 13) experienced a 0.69 year decrement. For individuals in relatively richer villages and for Han youth, we do not observe significant educational attainment decrements due to closure. We do find that the direction of closure effects are generally relatively more negative for girls, both from relatively richer and poorer villages.

In columns 3 and 4 of Table \ref{tab:interhan:bygender:byvilrich:attain:nsfto}, we find that closure has the strongest negative effects on Mandarin writing for relatively poorer minority girls as well as boys who were 6 to 9 at year of closure. The effects are also negative for minority girls who were 10 to 13 at year of closure. The magnitudes or the effects are similar to the results from Table \ref{tab:interhan:bygender:mandwritem2c:nsfto} in the combined sample. In the relatively poorer villages, we do not find significant effects of closure on Han individuals. In relatively richer villages, the effects of closure on Mandarin writing for girls are negative, but generally insignificant. The exception is the finding of significant negative effects for Han girls who were 10 to 13 years of age at the time of closure.

In this section, given the limited sample and the larger number of heterogeneous effects parameters, estimates are less precise and the chance of obtaining spuriously significant estimates is larger. Overall, we continue to find stronger negative effects of closure for minority youth compared to Han youth, and the negative effects are more concentrated in relatively poorer villages.
\section{Conclusion}
With a focus on regions of China with substantial minority populations, this paper has investigated whether there are different impacts of China's rural school consolidation initiative on educational attainment and written Mandarin capability for minority students, compared to Han students. Descriptive results showed, first, that consolidated schools were safer and had generally better equipment than village schools, but were farther away and more likely to require boarding, which might increase perceived risk and cost. Consolidated schools were also larger, though we are not able to observe whether classes were larger. These pull- and push- factors in consolidated schools are likely to affect minority and Han youth in a similar though not necessarily identical fashion. However, consolidated schools were reportedly less likely than remaining village schools to offer a minority language of instruction, and the differences were stark. Across all sample villages, 31 percent of villages with schools and 15 percent of villages with closure reported a minority language of instruction. In predominantly minority villages, 43 percent of villages with schools, but only 22 percent of closure villages, reported that the primary school offered minority language of instruction. The difference in language of instruction associated with consolidation would likely be relevant to minority but not Han youth--perhaps hindering early engagement and learning in ways that might translate to poorer educational foundations, and perhaps heightening the perceived risks associated with distance and boarding.

Beyond language mismatch, children from culturally distinct groups might also find themselves in classrooms in which they need to learn different behavioral norms. For example, \textcite{liu_measuring_2019} reported that when children from the matriarchal Mosuo ethnic group and Han students came together at school, ethnic differences in gendered patterns of risk-taking behavior observable at early ages diminished as Mosuo children adopted the risk preferences of the majority. We are unable to observe subtle differences in school, classroom, or peer culture in the current study, but these topics might be very relevant for understanding minoritized students' experiences of consolidation.

Our main analyses show clear evidence of an average negative effect of closure on two of the three outcomes studied, educational attainment and reported written Mandarin ability, for minority youth, compared to Han youth. Importantly, the average decrements experienced by minority youth masked significant heterogeneities. This paper is novel in attending to the possibility that minoritized status, together with gender and community economic status and ethnic composition,  might intersect to generate distinct vulnerabilities to policy impact.  Our analysis shows that the intersectional perspective does yield useful insights about heterogeneous policy impacts.  For example, for both educational attainment and written language facility, penalties accruing to minority youth occurred only in poorer villages and not to minority youth in wealthier villages. In addition, penalties were generally heavier for minority girls.  However, in the most ethnically segregated minority villages, boys from minority families were highly vulnerable to closure effects on educational attainment and written Mandarin facility. This latter pattern could emerge if boys were particularly affected by the linguistic mismatch potentially associated with consolidation in predominantly minority villages, but we do not have a means to test this speculation.

In an age of declining fertility and urbanization, sparse school-aged populations in rural communities are becoming more common. Press reports suggest that school consolidation initiatives are emerging in a variety of  settings as a policy response \autocite[for example, see][]{chowdhury_cramped_2017, malik_apprehensions_2013, harun_schools_2017, saengpassa_ministrys_2017, setiawati_schools_2010, tawie_sarawak_2017}.  China has been at the leading edge of this policy trend, with large-scale rural primary school consolidations occurring over the course of the early 2000s. While the likely impacts of school consolidations are very context-specific, the rationales across the world are strikingly similar, as are the concerns about potential implications for children in socially and economically marginalized communities.  Our findings speak to the importance of considering intersecting dimensions of stratification together, when assessing the policy's impacts.

\clearpage
\pagebreak

%%%%%%%%%%%%%%%%%%%%%%%%%%%%%%%%%%%
% Part III. Bibliography
%%%%%%%%%%%%%%%%%%%%%%%%%%%%%%%%%%%
\begingroup
\setstretch{1.0}
\setlength\bibitemsep{3pt}
\printbibliography[title=References]
\endgroup
\pagebreak

%%%%%%%%%%%%%%%%%%%%%%%%%%%%%%%%%%%%
% Part IV. Tables and figures
%%%%%%%%%%%%%%%%%%%%%%%%%%%%%%%%%%%%
\processdelayedfloats

%%%%%%%%%%%%%%%%%%%%%%%%%%%%%%%%%%%%
% PART V. Appendix
%%%%%%%%%%%%%%%%%%%%%%%%%%%%%%%%%%%%
\appendix

\makeatletter
\efloat@restorefloats
\makeatother

\setlength{\footnotemargin}{5.75mm}
% Title for Appendix
\begingroup
\doublespacing
\centering
\Large ONLINE APPENDIX \\
\Large\begin{singlespace}\href{\PAPERDOIURL}{\PAPERTITLE}\end{singlespace}
\large \AUTHORHANNUM{} and \AUTHORWANG{}\\[1.0em]
\endgroup
% \clearpage

% Online appendix
%%%%%%%%%%%%%%%%%%%%%%%%%%%%%%%%%%%%
% Appendix A, Data details
%%%%%%%%%%%%%%%%%%%%%%%%%%%%%%%%%%%%

% Set equation, figure, table indexing
\renewcommand{\thetable}{A.\arabic{table}}
\setcounter{table}{0}
\renewcommand{\thefootnote}{A.\arabic{footnote}}
\setcounter{footnote}{0}

\section{Additional data details\label{sec:amoredata}}
\subsection{Location of closures}
\label{sec:aloc}

This paper utilizes data from the rural sample of the China Household
Ethnic Survey (CHES 2011), which covers households and villages from 728 villages.\footnote{There are 751 unique village IDs in the survey, but 17 villages do not have school closure information, and 6 villages report closure without a closure year.}
Following discussions in Section \ref{sec:data:sample}, our analytic sample includes 638 villages across 7 provinces: Qinghai Province (107 villages);
Ningxia Hui Autonomous Region (88 villages); Xinjiang Uygur
Autonomous Region (80 villages); Inner Mongolia Autonomous Region (74 villages); Qiandongnan Miao and Dong Autonomous Prefecture in Guizhou
Province (110 villages); Hunan Province (87 villages); and Guangxi Zhuang Autonomous Region (92 villages).

\subsection{Sampling procedure}
\label{sec:asampling}

The CHES survey is the largest-scale cross-province survey ever gathered to study the socio-economic conditions of minorities in ethnically diverse regions of China. It was designed by China's Academy of Social Sciences and the Central Nationalities University, and it was administrated by local offices of the National Bureau of Statistics (NBS). The survey villages were selected based on a subset of the NBS's Rural Household Survey (RHS) \autocite{national_bureau_of_statistics_of_china_china_2012}. Selected villages are not representative of their respective provinces and autonomous regions overall, but are from prefectures with substantial minority populations in order for the survey to capture socio-economic conditions of minorities and Han individuals in ethnically diverse areas. Households in villages were selected by systematic sampling based on their agricultural census address codes. The survey was implemented in early 2012 and asked households to report information from the end of 2011. Household surveys are complemented by surveys of villages heads from the sampled villages. \textcite{gustafsson_ethnicity_2019} provide more information on the sampling procedure and other information related to the survey.

\subsection{Ethnic composition of the sample\label{sec:minobreakdown}}
The sample we analyze comes from seven provinces of China with large concentrations of minority populations. China is home to 56 minority nationalities, including the Han majority and 55 minority nationalities. In our analytic sample, the predominant minority group in each of the seven provinces differs. We report here minority groups that account for more than 3 percent of the overall analytic sample, including Han individuals, for each province: Tibetan, Hui, Salar, and Tu account for 31, 19, 9, and 3 percent of the analytical sample in Qinghai province; Hui account for 46 percent of the analytical sample in the Ningxia Hui Autonomous Region; Uygur, Kazaks, and Hui account for 58, 9, and 3 percent of the analytic sample in the Xinjiang Uygur Autonomous Region; Mongolians account for 23 percent of the analytic sample in the Inner Mongolia Autonomous Region; Miao and Dong account for 49 and 27 percent of the analytic sample from the Qiandongnan Miao and Dong Autonomous Prefecture in Guizhou Province; Miao, Tujia, Dong, and Yao account 34, 17, 16, and 6 percent of the analytical sample in Hunan province; and Zhuang, Yao, Miao, Dong, and Mulao account for 37, 9, 8, 5, and 4 percent of the analytical in the Guangxi Zhuang Autonomous Region.

There are 28 minority groups included in the analytical sample. The ten largest minority groups in our analytic sample are Miao, Hui, Uygur, Dong, Zhuang, Tibetan, Tujia, Mongolian, Yao, and Salar; they account for, collectively, 94 percent of the overall minority population within the analytic sample. These ten minority groups accounted for 66 percent of the overall minority population in China in 2010, and eight of these ten minority groups are among the largest ten minority groups in China overall \autocite{china_2020_census}. Manchus and Yi are the 3rd and 6th largest minority groups in China, and they jointly account for 17.1 percent of the minority population in China in 2010. The analytical sample has less than 0.1 percent Manchus and no Yi individuals.\footnote{Manchus largely reside in Northern and Northeastern China, generally in provinces with predominantly Han populations. The Yi people largely reside in Yunnan province, which is not a part of the survey sample we use in this study.}

In Section \ref{sec:app:esti:minoownlang}, we consider the robustness of our overall estimation results to the inclusion of minorities without their own language. 

\subsection{Age of enrollment, age at closure, and age in 2011\label{sec:app:stats:age}}

In this study, we define age cutoffs for the educational attainment based on primary school enrollment age patterns in our data from 2011. Overall enrollment rates into primary school jump from 11 percent at age 5 to 76 percent between ages 6 to 8. Enrollment rates decrease from 50 percent at age 13 to 15 percent at age 14. A lower proportion of 5 year old ethnic minorities are enrolled in primary schools (compared to Han children), and a lower proportion of 12 to 15 year old ethnic minorities are enrolled between ages 12 and 15 in middle school and beyond. Overall, children are predominantly enrolled in primary schools between 6 and 13 years of age in 2011. Hence, for villages with closure, we consider children who were between 6 and 13 years of age at the year of closure as most likely to have been enrolled in primary schools.

In Table \ref{tab:stats:indi:ageatclosure}, we present statistics on educational attainment and language facilities for individuals who experienced closure at different ages conditional on their cohorts (age in 2011). The columns are grouped by age in 2011, using five-year intervals between 10 to 34 years of age. Row Groups A, B, C and D show statistics for individuals experiencing closure when they were 6 to 9, 10 to 13, 14 to 21, and 22 to 29 years of age. Row Group E shows statistics for individuals from villages without closure.

\begin{table}[htbp]
\centering
\def\sym#1{\ifmmode^{#1}\else\(^{#1}\)\fi}
\caption{Educational attainment and Mandarin ability by cohort (2011 age) and age at closure\label{tab:stats:indi:ageatclosure}}
\begin{adjustbox}{max width=1\textwidth}
\begin{tabular}{m{7cm} >{\centering\arraybackslash}m{1.5cm} >{\centering\arraybackslash}m{1.5cm} >{\centering\arraybackslash}m{1.5cm} >{\centering\arraybackslash}m{1.5cm} >{\centering\arraybackslash}m{1.5cm}}
\toprule
&
\multicolumn{5}{c}{\small
Age
in
2011}
\\
\cmidrule(l{5pt}r{5pt}){2-6}
&
\multicolumn{1}{C{1.5cm}}{\textbf{\small
10-14}}
&
\multicolumn{1}{C{1.5cm}}{\textbf{\small
15-19}}
&
\multicolumn{1}{C{1.5cm}}{\textbf{\small
20-24}}
&
\multicolumn{1}{C{1.5cm}}{\textbf{\small
25-29}}
&
\multicolumn{1}{C{1.5cm}}{\textbf{\small
30-34}}
\\
\midrule
\multicolumn{6}{L{14.5cm}}{\vspace*{-5mm}\hspace*{-21mm}\textbf{{\normalsize Group A: Age 6 to 9 at year of closure}}} \\&            &            &            &            &            \\
Age in 2011 (mean)  &       11.85&       16.84&       20.20&            &            \\
Years of education (mean)&        5.46&        9.76&       10.93&            &            \\
Strong spoken Mandarin fraction&        0.23&        0.62&        0.53&            &            \\
Strong written Mandarin fraction&        0.21&        0.41&        0.53&            &            \\
\midrule
Observations        &         169&          58&          15&           &           \\

\midrule
\multicolumn{6}{L{14.5cm}}{\vspace*{-5mm}\hspace*{-21mm}\textbf{{\normalsize Group B: Age 10 to 13 at year of closure}}} \\&            &            &            &            &            \\
Age in 2011 (mean)  &       12.59&       16.77&       21.53&       25.00&            \\
Years of education (mean)&        5.91&        9.38&       10.38&        8.50&            \\
Strong spoken Mandarin fraction&        0.23&        0.44&        0.55&        1.00&            \\
Strong written Mandarin fraction&        0.17&        0.38&        0.45&        1.00&            \\
\midrule
Observations        &         129&         222&         113&           2&           \\

\midrule
\multicolumn{6}{L{14.5cm}}{\vspace*{-5mm}\hspace*{-21mm}\textbf{{\normalsize Group C: Age 14 to 21 at year of closure}}} \\&            &            &            &            &            \\
Age in 2011 (mean)  &       14.00&       17.76&       21.91&       26.26&       30.74\\
Years of education (mean)&        7.00&        9.72&       10.23&        8.96&        8.41\\
Strong spoken Mandarin fraction&        0.38&        0.42&        0.42&        0.42&        0.41\\
Strong written Mandarin fraction&        0.19&        0.40&        0.37&        0.34&        0.30\\
\midrule
Observations        &          16&         275&         581&         224&          27\\

\midrule
\multicolumn{6}{L{14.5cm}}{\vspace*{-5mm}\hspace*{-21mm}\textbf{{\normalsize Group D: Age 22 to 29 at year of closure}}} \\&            &            &            &            &            \\
Age in 2011 (mean)  &            &            &       23.26&       27.18&       31.72\\
Years of education (mean)&            &            &        9.98&        8.75&        8.09\\
Strong spoken Mandarin fraction&            &            &        0.31&        0.28&        0.33\\
Strong written Mandarin fraction&            &            &        0.21&        0.23&        0.24\\
\midrule
Observations        &           &           &         101&         452&         314\\

\midrule
\multicolumn{6}{L{14.5cm}}{\vspace*{-5mm}\hspace*{-21mm}\textbf{{\normalsize Group E: Non-closure villages individuals}}} \\&            &            &            &            &            \\
Age in 2011 (mean)  &       12.03&       17.06&       22.00&       26.89&       31.82\\
Years of education (mean)&        5.48&        9.30&        9.79&        8.51&        7.47\\
Strong spoken Mandarin fraction&        0.19&        0.31&        0.33&        0.25&        0.20\\
Strong written Mandarin fraction&        0.15&        0.32&        0.28&        0.18&        0.14\\
\midrule
Observations        &        1565&        1837&        2343&        1747&        1447\\
\bottomrule
\addlinespace[-1.5em]
\multicolumn{6}{L{16.55cm}}{
\footnotesize
\justify
Group A individuals were 6 to 9 at the year of closure and were exposed to consolidated primary school for up to half of their primary school years. Group B individuals were 10 to 13 at the year of closure and transitioned from village schools to consolidated primary schools during the final years of primary school. Group C and D individuals are from villages that experienced closure but were beyond primary school age at the year of closure. Group E individuals are from villages without school closure.
}\\
\end{tabular}
\end{adjustbox}
\end{table}
\newpage

\subsection{Years of education and Mandarin ability by ethnic category and closure\label{sec:app:stats:attain}}
In Table \ref{tab:stats:indi:fourgrps}, we present individual-level summary statistics by minority and closure status. Across 5-year interval cohort groups, Table \ref{tab:stats:indi:fourgrps} shows that Han and minority individuals in closure villages have higher years of education completed and larger fractions reporting strong Mandarin abilities.

The raw differences in years of education and strong Mandarin ability between closure and non-closure villages is largely explained by province fixed effects, age fixed effects, and gender. Specifically, we consider a sample of individuals who were over 20 years of age in 2011 and include only individuals who were older than 13 at the year of closure---this group includes individuals in columns 3, 4 and 5 and Groups C, D and E from Table \ref{tab:stats:indi:ageatclosure}. Using this sample, we regress years of education on village closure status, controlling for province fixed effects, cohort (age in 2011) fixed effects and gender. The coefficient for closure on years of education is 0.09 (s.e. 0.10) and insignificant. We interpret this as a measure of the existing difference in educational attainment between closure and non-closure villages in the absence of school closure. Interacting closure with Han status, results remain insignificant: we find that among Han individuals, the coefficient for closure is -0.20 (s.e. 0.19), and among minority individuals, the coefficient for closure is 0.17 (s.e. 0.12).

With the same sample and controls, we also regress strong written Mandarin ability on closure. The coefficient from linear probability regression for closure on reporting strong Mandarin ability is 0.025 (s.e. 0.013). Interacting closure with Han status, we find that among Han individuals, the coefficient for closure is -0.015 (s.e. 0.025), and among minority individuals, the coefficient for closure is 0.030 (s.e. 0.016).

\begin{table}[htbp]
\centering
\def\sym#1{\ifmmode^{#1}\else\(^{#1}\)\fi}
\caption{Means of individual-level variables by cohort (2011 age), ethnic category and closure\label{tab:stats:indi:fourgrps}}
\begin{adjustbox}{max width=1\textwidth}
\begin{tabular}{m{7cm} >{\centering\arraybackslash}m{1.5cm} >{\centering\arraybackslash}m{1.5cm} >{\centering\arraybackslash}m{1.5cm} >{\centering\arraybackslash}m{1.5cm} >{\centering\arraybackslash}m{1.5cm}}
\toprule
&
\multicolumn{5}{c}{\small
Age
in
2011}
\\
\cmidrule(l{5pt}r{5pt}){2-6}
&
\multicolumn{1}{C{1.5cm}}{\textbf{\small
10-14}}
&
\multicolumn{1}{C{1.5cm}}{\textbf{\small
15-19}}
&
\multicolumn{1}{C{1.5cm}}{\textbf{\small
20-24}}
&
\multicolumn{1}{C{1.5cm}}{\textbf{\small
25-29}}
&
\multicolumn{1}{C{1.5cm}}{\textbf{\small
30-34}}
\\
\midrule
\multicolumn{6}{L{14.5cm}}{\vspace*{-5mm}\hspace*{-21mm}\textbf{{\normalsize Panel A: Minority in closure villages}}} \\&            &            &            &            &            \\
Male fraction       &        0.51&        0.58&        0.51&        0.55&        0.60\\
Age 6-9 at year of closure fraction&        0.50&        0.11&        0.00&        0.00&        0.00\\
Age 6-13 at year of closure fraction&        0.96&        0.46&        0.12&        0.00&        0.00\\
Currently enrolled fraction&        0.90&        0.55&        0.11&        0.02&        0.01\\
Years of education  &        5.47&        9.12&        9.68&        8.60&        8.01\\
Strong spoken Mandarin fraction&        0.19&        0.38&        0.34&        0.26&        0.23\\
Strong written Mandarin fraction&        0.14&        0.32&        0.27&        0.22&        0.22\\
\midrule
Observations        &         175&         309&         468&         445&         183\\

\midrule
\multicolumn{6}{L{14.5cm}}{\vspace*{-5mm}\hspace*{-21mm}\textbf{{\normalsize Panel B: Han in closure villages}}} \\&            &            &            &            &            \\
Male fraction       &        0.42&        0.52&        0.55&        0.55&        0.51\\
Age 6-9 at year of closure fraction&        0.58&        0.10&        0.04&        0.00&        0.00\\
Age 6-13 at year of closure fraction&        0.94&        0.57&        0.21&        0.01&        0.00\\
Currently enrolled fraction&        0.88&        0.59&        0.21&        0.00&        0.00\\
Years of education  &        6.04&       10.17&       11.00&        9.23&        8.25\\
Strong spoken Mandarin fraction&        0.29&        0.54&        0.56&        0.45&        0.46\\
Strong written Mandarin fraction&        0.27&        0.48&        0.49&        0.36&        0.27\\
\midrule
Observations        &         139&         246&         342&         233&         158\\

\midrule
\multicolumn{6}{L{14.5cm}}{\vspace*{-5mm}\hspace*{-21mm}\textbf{{\normalsize Panel C: Minority in non-closure villages}}} \\&            &            &            &            &            \\
Male fraction       &        0.53&        0.51&        0.52&        0.55&        0.55\\
Currently enrolled fraction&        0.93&        0.52&        0.13&        0.01&        0.00\\
Years of education  &        5.41&        9.05&        9.39&        8.24&        7.22\\
Strong spoken Mandarin fraction&        0.16&        0.26&        0.27&        0.21&        0.16\\
Strong written Mandarin fraction&        0.14&        0.27&        0.22&        0.14&        0.11\\
\midrule
Observations        &        1133&        1285&        1619&        1230&        1026\\

\midrule
\multicolumn{6}{L{14.5cm}}{\vspace*{-5mm}\hspace*{-21mm}\textbf{{\normalsize Panel D: Han in non-closure villages}}} \\&            &            &            &            &            \\
Male fraction       &        0.50&        0.55&        0.50&        0.53&        0.53\\
Currently enrolled fraction&        0.95&        0.66&        0.21&        0.03&        0.01\\
Years of education  &        5.67&        9.89&       10.68&        9.16&        8.08\\
Strong spoken Mandarin fraction&        0.26&        0.44&        0.46&        0.34&        0.30\\
Strong written Mandarin fraction&        0.20&        0.43&        0.41&        0.25&        0.20\\
\midrule
Observations        &         432&         552&         724&         517&         421\\
\bottomrule
\addlinespace[-1.5em]
\multicolumn{6}{L{16.55cm}}{
\footnotesize
\justify
Means of individual-level variables across cohorts. The analytic sample includes individuals between 10 to 34 years of age in 2011. In villages with closure, individuals are included if they were also 6 to 29 years of age at the village-specific years of closure.
}\\
\end{tabular}
\end{adjustbox}
\end{table}
  \clearpage
\pagebreak

\subsection{Written Mandarin ability responses\label{sec:app:stats:mandwrite}}
In Table \ref{tab:stats:indi:mandwritegroups}, we summarize enrollment, educational attainment, and share reporting strong spoken Mandarin ability by categories of the household respondent-provided rating of written Mandarin ability.

\begin{table}[htbp]
	\centering
	\def\sym#1{\ifmmode^{#1}\else\(^{#1}\)\fi}
	\caption{Enrollment, attainment and spoken Mandarin ability by written Mandarin ability\label{tab:stats:indi:mandwritegroups}}
	\begin{adjustbox}{max width=1\textwidth}
		\begin{tabular}{m{7cm} >{\centering\arraybackslash}m{1.5cm} >{\centering\arraybackslash}m{1.5cm} >{\centering\arraybackslash}m{1.5cm} >{\centering\arraybackslash}m{1.5cm} >{\centering\arraybackslash}m{1.5cm}}
			\toprule
			&
			\multicolumn{5}{c}{\small
				Age
				in
				2011}
			\\
			\cmidrule(l{5pt}r{5pt}){2-6}
			&
			\multicolumn{1}{C{1.5cm}}{\textbf{\small
					10-14}}
			&
			\multicolumn{1}{C{1.5cm}}{\textbf{\small
					15-19}}
			&
			\multicolumn{1}{C{1.5cm}}{\textbf{\small
					20-24}}
			&
			\multicolumn{1}{C{1.5cm}}{\textbf{\small
					25-29}}
			&
			\multicolumn{1}{C{1.5cm}}{\textbf{\small
					30-34}}
			\\
			\midrule
			\multicolumn{6}{L{14.5cm}}{\vspace*{-5mm}\hspace*{-21mm}\textbf{{\normalsize Panel A: Non-response for written Mandarin ability}}} \\&            &            &            &            &            \\
			Currently enrolled fraction&        0.93&        0.50&        0.12&        0.01&        0.01\\
			Years of education  &        5.52&        9.25&        9.60&        8.54&        7.50\\
			Strong spoken Mandarin fraction&        0.01&        0.01&        0.01&        0.04&        0.01\\
			\midrule
			Observations        &         260&         287&         277&         181&         137\\

			\midrule
			\multicolumn{6}{L{14.5cm}}{\vspace*{-5mm}\hspace*{-21mm}\textbf{{\normalsize Panel B: No written Mandarin ability}}} \\&            &            &            &            &            \\
			Currently enrolled fraction&        0.87&        0.33&        0.00&        0.00&        0.01\\
			Years of education  &        4.58&        7.13&        6.15&        4.98&        3.79\\
			Strong spoken Mandarin fraction&        0.01&        0.04&        0.04&        0.04&        0.03\\
			\midrule
			Observations        &         101&          91&         205&         212&         203\\

			\midrule
			\multicolumn{6}{L{14.5cm}}{\vspace*{-5mm}\hspace*{-21mm}\textbf{{\normalsize Panel C: Simple written Mandarin ability}}} \\&            &            &            &            &            \\
			Currently enrolled fraction&        0.92&        0.41&        0.03&        0.01&        0.00\\
			Years of education  &        4.84&        8.19&        8.12&        7.38&        6.82\\
			Strong spoken Mandarin fraction&        0.04&        0.05&        0.05&        0.07&        0.07\\
			\midrule
			Observations        &         478&         274&         435&         506&         421\\

			\midrule
			\multicolumn{6}{L{14.5cm}}{\vspace*{-5mm}\hspace*{-21mm}\textbf{{\normalsize Panel D: Good written Mandarin ability}}} \\&            &            &            &            &            \\
			Currently enrolled fraction&        0.94&        0.53&        0.07&        0.01&        0.00\\
			Years of education  &        5.82&        9.20&        9.70&        9.00&        8.38\\
			Strong spoken Mandarin fraction&        0.16&        0.16&        0.21&        0.18&        0.18\\
			\midrule
			Observations        &         738&         937&        1282&        1037&         743\\

			\midrule
			\multicolumn{6}{L{14.5cm}}{\vspace*{-5mm}\hspace*{-21mm}\textbf{{\normalsize Panel E: Strong written Mandarin ability}}} \\&            &            &            &            &            \\
			Currently enrolled fraction&        0.93&        0.71&        0.35&        0.04&        0.00\\
			Years of education  &        6.17&       10.26&       11.90&       10.60&        9.44\\
			Strong spoken Mandarin fraction&        0.78&        0.81&        0.86&        0.85&        0.82\\
			\midrule
			Observations        &         302&         803&         954&         489&         284\\
			\bottomrule
			\addlinespace[-1.5em]
			\multicolumn{6}{L{16.55cm}}{
				\footnotesize
				\justify
				The Mandarin written ability variable allows for four answers: no, simple, good, and strong. The answer is self-rated and provided by the household respondent. Approximately 10 percent of the respondents of the analytical sample did not answer this question. Better Mandarin written ability positively relates enrollment, educational attainment, and better spoken Mandarin ability.
			}\\
		\end{tabular}
	\end{adjustbox}
\end{table}
\clearpage

%%%%%%%%%%%%%%%%%%%%%%%%%%%%%%%%%%%%
% Appendix B, Solution and Estimation Details
%%%%%%%%%%%%%%%%%%%%%%%%%%%%%%%%%%%%
\section{Additional estimation results}

% Set equation, figure, table indexing
\renewcommand{\thefigure}{B.\arabic{figure}}
\setcounter{figure}{0}
\renewcommand{\thetable}{B.\arabic{table}}
\setcounter{table}{0}
\renewcommand{\theequation}{B.\arabic{equation}}
\setcounter{equation}{0}
\renewcommand{\thefootnote}{B.\arabic{footnote}}
\setcounter{footnote}{0}

\subsection{Minority groups without own language\label{sec:app:esti:minoownlang}}
\newcommand{\tablenoteexcludeone}{\,Hui, Miao, Yao, Dong, and Tujia minorities account for 59, 13, 8, 6, and 5 percent of the excluded no-own-language individuals.}
\newcommand{\tablenoteexcludetwo}{\,Excluding Hui, Dongxiang, Lisu and Manchu individuals. 95 percent of excluded are Hui.}
There is significant linguistic and cultural diversity among and within ethnic groups \autocite{harrell_linguistics_1993, dwyer_texture_1998}. While most minority groups in China have at least one spoken language, some do not \autocite{wang_chinas_2015}. The linguistic distinctions might capture not only differences in language usage but also other aspects of culture, all of which might relate to the effects of closure. For robustness, in this section, we re-estimate our main educational attainment and Mandarin ability specifications from Tables \ref{tab:interhan:attain:nsfto} and
\ref{tab:interhan:mandwritetalkm2c:nsfto}
but now exclude minority individuals that do not have their own language. Remaining minority individuals are potentially more sharply distinct from Han individuals.

The CHES survey asks, for each individual, whether household member have their own language.\footnote{The Chinese survey question was: \chinese{是否有本民族语言}, shi fou you ben min zu yu yan.} Among the minority respondents, 84 percent of individuals were reported as ``\emph{yes}'' on this question. Individuals from Hui, Miao, Yao, Dong, and Tujia minorities account for 59, 13, 8, 6, and 5 percent of the total number of \emph{no} respondents, respectively.\footnote{Looking within each minority group, 60, 10, 41, 8, and 22 percent of Hui, Miao, Yao, Dong, and Tujia individuals were reported as ``\emph{no}'' on the question, respectively.} Among minorities with more than three respondents in the survey, four ethnic groups had a higher than 50 percent \emph{no} share: Hui (60 percent among 1213 respondents),\footnote{In the analytic sample, 71 and 25 percent of the Hui respondents come from Ningxia and Qinghai, respectively. 52 and 79 percent of Ningxia and Qinghai Hui individuals reported \emph{no} to the having own minority language question.} Dongxiang (81 percent among 36 respondents), Lisu (94 percent among 17 respondents), and Manchu (80 percent among 5 respondents). Variations in responses might be due to language heterogeneity within ethnic groups \autocite{roche_articulating_2019} or possible differences in the interpretation of linguistic differences within minority groups.

In our first robustness exercise, we drop cases in which minority respondents are reported as not having a minority language. In our second robustness exercise, we drop cases for minority groups in which over 50\% of respondents are reported as not having a minority language. Hui individuals account for 59 and 95 percent of the excluded individuals under the first and second exclusion scenarios, respectively. In both exercises, the Han group stays the same. Attainment results under the first and second exclusion scenario are shown in columns 1 to 3 and 4 to 6 of Table \ref{tab:interhan:attain:nsfto:drophui}, respectively. Table \ref{tab:interhan:mandwritetalkm2c:nsfto:drophui} presents results for Mandarin ability following the same layout. In all cases, we find that the closure effects coefficients are very similar in the direction and magnitude to the coefficients shown in tables \ref{tab:interhan:attain:nsfto} and
\ref{tab:interhan:mandwritetalkm2c:nsfto}.\footnote{In an additional set of exercises, we use all data and estimate heterogeneous effects for Han individuals, minorities with own languages, and minorities without own languages. Estimates for minorities with and without own language have similar directions, but estimates for minorities without own language have standard errors that are more than twice as large due to the limited sample size. Results are available upon request from the authors, but not presented here for conciseness.}

The robustness of our main estimation results to considering only minorities with own language is not surprising given that minorities without-own-language account for a small proportion of the overall minority sample.  These robustness checks do  indicate that the inclusion of these individuals does not drive the overall results. Given limited sample size for the without own language minority groups exposed to closure in the relevant age-at-closure year ranges, we are unable to obtain reliable estimates of whether Hui and other individuals are impacted more or less by closure, compared to more linguistically distinct minority groups.

\begin{table}[htbp]
\centering
\caption{Regressions of educational attainment with school closure, Han-minority interaction, excluding minorities without own language\label{tab:interhan:attain:nsfto:drophui}}
\begin{adjustbox}{max width=1\textwidth}
\begin{tabular}{m{8.5cm} >{\centering\arraybackslash}m{1.5cm} >{\centering\arraybackslash}m{1.5cm} >{\centering\arraybackslash}m{1.5cm}
>{\centering\arraybackslash}m{1.5cm} >{\centering\arraybackslash}m{1.5cm} >{\centering\arraybackslash}m{1.5cm}}
\toprule
&
\multicolumn{6}{c}{
Outcome:
grades
completed
by
year
2011}
\\
\cmidrule(l{5pt}r{5pt}){2-7}
&
\multicolumn{3}{c}{\shortstack{Drop minority respondents\\reported as not having\\a minority language\,\dag}}
&
\multicolumn{3}{c}{\shortstack{Drop minority groups with\\$>50$\% respondents reported\\as not having own language.\,\ddag}}
\\
\cmidrule(l{5pt}r{5pt}){2-4}
\cmidrule(l{5pt}r{5pt}){5-7}
&
\multicolumn{1}{C{1.5cm}}{All villages}
&
\multicolumn{1}{C{1.5cm}}{Han $<90$\%}
&
\multicolumn{1}{C{1.5cm}}{Han $<70$\%}
&
\multicolumn{1}{C{1.5cm}}{All villages}
&
\multicolumn{1}{C{1.5cm}}{Han $<90$\%}
&
\multicolumn{1}{C{1.5cm}}{Han $<70$\%}
\\
\midrule
\multicolumn{7}{L{19cm}}{\vspace*{-3.5mm}\hspace*{-8mm}\textit{Baseline group: Child was 14-21 years old at village primary school closure year}} \\
\multicolumn{7}{L{19cm}}{\vspace*{-5mm}\hspace*{-8mm}{Child is minority interaction:}} \\
& & & & & & \\
\vspace*{0mm}\hspace*{5mm} Closure $\times$ age at closure 6-9
& -0.51\sym{**} & -0.52\sym{**} & -0.43\sym{*} & -0.57\sym{**} & -0.59\sym{**} &  -0.51\sym{**} \\
&\vspace*{-2mm}{\footnotesize (0.25)} &\vspace*{-2mm}{\footnotesize (0.25)} &\vspace*{-2mm}{\footnotesize (0.26)} &\vspace*{-2mm}{\footnotesize (0.24)} &\vspace*{-2mm}{\footnotesize (0.25)} &\vspace*{-2mm}{\footnotesize (0.25)}\\
\vspace*{0mm}\hspace*{5mm} Closure $\times$ age at closure 10-13
& -0.51\sym{***}& -0.56\sym{***}& -0.47\sym{***}& -0.56\sym{***}& -0.60\sym{***} & -0.50\sym{**}\\
&\vspace*{-2mm}{\footnotesize (0.19)} &\vspace*{-2mm}{\footnotesize (0.20)} &\vspace*{-2mm}{\footnotesize (0.20)} &\vspace*{-2mm}{\footnotesize (0.20)} &\vspace*{-2mm}{\footnotesize (0.20)} &\vspace*{-2mm}{\footnotesize (0.20)}\\
\vspace*{0mm}\hspace*{5mm} Closure $\times$ age at closure 22-29
& 0.19 & 0.14 & 0.22 & 0.11 & 0.072 & 0.14\\
&\vspace*{-2mm}{\footnotesize (0.18)} &\vspace*{-2mm}{\footnotesize (0.19)} &\vspace*{-2mm}{\footnotesize (0.19)} &\vspace*{-2mm}{\footnotesize (0.18)} &\vspace*{-2mm}{\footnotesize (0.19)} &\vspace*{-2mm}{\footnotesize (0.20)}\\
\multicolumn{7}{L{19cm}}{\vspace*{-5mm}\hspace*{-8mm}{Child is Han interaction:}} \\
& & & & & & \\
\vspace*{0mm}\hspace*{5mm} Closure $\times$ age at closure 6-9
& -0.024 & -0.53 & -0.13 & -0.051 & -0.41 & 0.15   \\
&\vspace*{-2mm}{\footnotesize (0.26)} &\vspace*{-2mm}{\footnotesize (0.39)} &\vspace*{-2mm}{\footnotesize (0.39)} &\vspace*{-2mm}{\footnotesize (0.26)} &\vspace*{-2mm}{\footnotesize (0.39)} &\vspace*{-2mm}{\footnotesize (0.37)} \\
\vspace*{0mm}\hspace*{5mm} Closure $\times$ age at closure 10-13
& -0.32 & -0.31 & -0.021 & -0.25 & -0.21 & 0.12   \\
&\vspace*{-2mm}{\footnotesize (0.20)} &\vspace*{-2mm}{\footnotesize (0.32)} &\vspace*{-2mm}{\footnotesize (0.41)} &\vspace*{-2mm}{\footnotesize (0.20)} &\vspace*{-2mm}{\footnotesize (0.32)} &\vspace*{-2mm}{\footnotesize (0.40)} \\
\vspace*{0mm}\hspace*{5mm} Closure $\times$ age at closure 22-29
& -0.24 & -0.19 & -0.15 & -0.14 & -0.024 & 0.046   \\
&\vspace*{-2mm}{\footnotesize (0.26)} &\vspace*{-2mm}{\footnotesize (0.33)} &\vspace*{-2mm}{\footnotesize (0.40)} &\vspace*{-2mm}{\footnotesize (0.25)}  &\vspace*{-2mm}{\footnotesize (0.32)} &\vspace*{-2mm}{\footnotesize (0.38)} \\
\midrule
\hspace*{3mm}Observations & 10386 & 8035 & 7251 & 10351 & 7973 & 7180   \\
\bottomrule
\addlinespace[-1.5em]
\multicolumn{7}{L{19.7cm}}{
\footnotesize\justify
\dag\tablenoteexcludeone\\
\ddag\tablenoteexcludetwo\\
Statistical significance:* 0.10 ** 0.05 *** 0.01. Robust standard error clustered at village level. Each column represents a separate regression. All regressions include village fixed effects, province-specific age FEs, Han-specific age FEs, and control for household size. Columns 2 to 5 sequentially drop villages with more than 90, 70, 50, and 10 percent Han population. Samples for columns 5 and 6 include only minority and Han individuals, respectively.
}\\
\end{tabular}
\end{adjustbox}
\end{table}
 \begin{table}[htbp]
\centering
\caption{Regressions of Mandarin ability with school closure, Han-minority status interaction, excluding minorities without own language\label{tab:interhan:mandwritetalkm2c:nsfto:drophui}}
\begin{adjustbox}{max width=1\textwidth}
\begin{tabular}{m{8.5cm} >{\centering\arraybackslash}m{1.5cm} >{\centering\arraybackslash}m{1.5cm} >{\centering\arraybackslash}m{1.5cm} >{\centering\arraybackslash}m{1.5cm} >{\centering\arraybackslash}m{1.5cm} >{\centering\arraybackslash}m{1.5cm}}
\toprule
&
\multicolumn{6}{c}{Outcome:
strong
Mandarin
ability
=
1,
no/simple/good
=
0}
\\
\cmidrule(l{5pt}r{5pt}){2-7}
&
\multicolumn{3}{c}{\shortstack{Drop minority respondents\\reported as not having\\a minority language\,\dag}}
&
\multicolumn{3}{c}{\shortstack{Drop minority groups with\\$>50$\% respondents reported\\as not having own language.\,\ddag}}
\\
\cmidrule(l{5pt}r{5pt}){2-4}
\cmidrule(l{5pt}r{5pt}){5-7}
&
\multicolumn{1}{C{1.5cm}}{All villages}
&
\multicolumn{1}{C{1.5cm}}{Han $<90$\%}
&
\multicolumn{1}{C{1.5cm}}{Han $<70$\%}
&
\multicolumn{1}{C{1.5cm}}{All villages}
&
\multicolumn{1}{C{1.5cm}}{Han $<90$\%}
&
\multicolumn{1}{C{1.5cm}}{Han $<70$\%}
\\
\midrule
\addlinespace[+1em]
\multicolumn{7}{L{19cm}}{\vspace*{1mm}\hspace*{-8mm}\textbf{Panel A: Written Mandarin ability}} \\
\multicolumn{7}{L{19cm}}{\vspace*{-3.5mm}\hspace*{-8mm}\textit{Baseline group: Child was 14-21 years old at village primary school closure year}} \\
\multicolumn{7}{L{19cm}}{\vspace*{-5mm}\hspace*{-8mm}{Child is minority interaction:}} \\
& & & & & &\\
\vspace*{0mm}\hspace*{5mm} Closure $\times$ age at closure 6-9
& -0.13\sym{***} & -0.15\sym{***} & -0.17\sym{***}
& -0.15\sym{***} & -0.15\sym{***} & -0.17\sym{***}\\
& \vspace*{-2mm}{\footnotesize (0.044)} & \vspace*{-2mm}{\footnotesize (0.046)} & \vspace*{-2mm}{\footnotesize (0.048)}
& \vspace*{-2mm}{\footnotesize (0.047)} & \vspace*{-2mm}{\footnotesize (0.048)} & \vspace*{-2mm}{\footnotesize (0.050)}\\
\vspace*{0mm}\hspace*{5mm} Closure $\times$ age at closure 10-13
& -0.069\sym{**} & -0.082\sym{**} & -0.083\sym{**}
& -0.062\sym{*} & -0.070\sym{**} & -0.070\sym{**}\\
& \vspace*{-2mm}{\footnotesize (0.034)} & \vspace*{-2mm}{\footnotesize (0.034)} & \vspace*{-2mm}{\footnotesize (0.034)}
& \vspace*{-2mm}{\footnotesize (0.033)} & \vspace*{-2mm}{\footnotesize (0.034)} & \vspace*{-2mm}{\footnotesize (0.034)}\\
\vspace*{0mm}\hspace*{5mm} Closure $\times$ age at closure 22-29
& -0.019 & -0.026 & -0.020
& -0.032 & -0.036 & -0.038\\
& \vspace*{-2mm}{\footnotesize (0.036)} & \vspace*{-2mm}{\footnotesize (0.036)} & \vspace*{-2mm}{\footnotesize (0.038)}
& \vspace*{-2mm}{\footnotesize (0.034)} & \vspace*{-2mm}{\footnotesize (0.035)} & \vspace*{-2mm}{\footnotesize (0.037)}\\
\multicolumn{7}{L{19cm}}{\vspace*{-5mm}\hspace*{-8mm}{Child is Han interaction:}} \\
& & & & & &\\
\vspace*{0mm}\hspace*{5mm} Closure $\times$ age at closure 6-9
& 0.061 & -0.067 & -0.083
& 0.068 & -0.047 & -0.038\\
& \vspace*{-2mm}{\footnotesize (0.055)} & \vspace*{-2mm}{\footnotesize (0.072)} & \vspace*{-2mm}{\footnotesize (0.092)}
& \vspace*{-2mm}{\footnotesize (0.055)} & \vspace*{-2mm}{\footnotesize (0.075)} & \vspace*{-2mm}{\footnotesize (0.10)}\\
\vspace*{0mm}\hspace*{5mm} Closure $\times$ age at closure 10-13
& -0.042 & -0.039 & -0.041
& -0.040 & -0.025 & -0.019\\
& \vspace*{-2mm}{\footnotesize (0.042)} & \vspace*{-2mm}{\footnotesize (0.060)} & \vspace*{-2mm}{\footnotesize (0.091)}
& \vspace*{-2mm}{\footnotesize (0.043)} & \vspace*{-2mm}{\footnotesize (0.062)} & \vspace*{-2mm}{\footnotesize (0.091)}\\
\vspace*{0mm}\hspace*{5mm} Closure $\times$ age at closure 22-29
& -0.029 & {\small -0.00032} & -0.072
& -0.021 & 0.014 & -0.053\\
& \vspace*{-2mm}{\footnotesize (0.039)} & \vspace*{-2mm}{\footnotesize (0.046)} & \vspace*{-2mm}{\footnotesize (0.053)}
& \vspace*{-2mm}{\footnotesize (0.038)} & \vspace*{-2mm}{\footnotesize (0.046)} & \vspace*{-2mm}{\footnotesize (0.049)}\\
\midrule
\hspace*{3mm}Observations & 9284 & 7002 & 6241 &  9429 &  7120 &  6346\\

\midrule
\addlinespace[+1em]
\multicolumn{7}{L{19cm}}{\vspace*{1mm}\hspace*{-8mm}\textbf{Panel B: Spoken Mandarin ability}} \\
\multicolumn{7}{L{19cm}}{\vspace*{-3.5mm}\hspace*{-8mm}\textit{Baseline group: Child was 14-21 years old at village primary school closure year}} \\
\multicolumn{7}{L{19cm}}{\vspace*{-5mm}\hspace*{-8mm}{Child is minority interaction:}} \\
& & & & & &\\
\vspace*{0mm}\hspace*{5mm} Closure $\times$ age at closure 6-9
& 0.0044 & 0.00077 & 0.0014
& -0.0090 & -0.0056 & -0.0078\\
& \vspace*{-2mm}{\footnotesize (0.054)} & \vspace*{-2mm}{\footnotesize (0.056)} & \vspace*{-2mm}{\footnotesize (0.058)}
& \vspace*{-2mm}{\footnotesize (0.053)} & \vspace*{-2mm}{\footnotesize (0.055)} & \vspace*{-2mm}{\footnotesize (0.057)}\\
\vspace*{0mm}\hspace*{5mm} Closure $\times$ age at closure 10-13
& -0.014 & -0.025 & -0.022
& -0.011 & -0.020 & -0.017\\
& \vspace*{-2mm}{\footnotesize (0.035)} & \vspace*{-2mm}{\footnotesize (0.036)} & \vspace*{-2mm}{\footnotesize (0.037)}
& \vspace*{-2mm}{\footnotesize (0.035)} & \vspace*{-2mm}{\footnotesize (0.036)} & \vspace*{-2mm}{\footnotesize (0.038)}\\
\vspace*{0mm}\hspace*{5mm} Closure $\times$ age at closure 22-29
& -0.0059 & -0.0068 & -0.0059
& -0.0047 & -0.0016 & -0.0010\\
& \vspace*{-2mm}{\footnotesize (0.034)} & \vspace*{-2mm}{\footnotesize (0.035)} & \vspace*{-2mm}{\footnotesize (0.035)}
& \vspace*{-2mm}{\footnotesize (0.032)} & \vspace*{-2mm}{\footnotesize (0.033)} & \vspace*{-2mm}{\footnotesize (0.035)}\\
\multicolumn{7}{L{19cm}}{\vspace*{-5mm}\hspace*{-8mm}{Child is Han interaction:}} \\
& & & & & &\\
\vspace*{0mm}\hspace*{5mm} Closure $\times$ age at closure 6-9
& 0.048 & -0.054 & 0.015
& 0.047 & -0.039 & 0.030\\
& \vspace*{-2mm}{\footnotesize (0.052)} & \vspace*{-2mm}{\footnotesize (0.076)} & \vspace*{-2mm}{\footnotesize (0.073)}
& \vspace*{-2mm}{\footnotesize (0.098)} & \vspace*{-2mm}{\footnotesize (0.070)} & \vspace*{-2mm}{\footnotesize (0.070)}\\
\vspace*{0mm}\hspace*{5mm} Closure $\times$ age at closure 10-13
& -0.0039 & -0.048 & 0.041
& -0.0017 & -0.028 & 0.059\\
& \vspace*{-2mm}{\footnotesize (0.036)} & \vspace*{-2mm}{\footnotesize (0.057)} & \vspace*{-2mm}{\footnotesize (0.068)}
& \vspace*{-2mm}{\footnotesize (0.036)} & \vspace*{-2mm}{\footnotesize (0.059)} & \vspace*{-2mm}{\footnotesize (0.066)}\\
\vspace*{0mm}\hspace*{5mm} Closure $\times$ age at closure 22-29
& -0.0089 & 0.014 & -0.016
& -0.0050 & 0.026 & -0.0050\\
& \vspace*{-2mm}{\footnotesize (0.041)} & \vspace*{-2mm}{\footnotesize (0.057)} & \vspace*{-2mm}{\footnotesize (0.070)}
& \vspace*{-2mm}{\footnotesize (0.041)} & \vspace*{-2mm}{\footnotesize (0.057)} & \vspace*{-2mm}{\footnotesize (0.067)}\\
\midrule
\hspace*{3mm}Observations & 9271 &  6998 &  6241 &  9420 &  7120 &  6350\\
\bottomrule
\addlinespace[-1.5em]
\multicolumn{7}{L{19.7cm}}{
\footnotesize\justify
\dag\tablenoteexcludeone\\
\ddag\tablenoteexcludetwo\\
Statistical significance:* 0.10 ** 0.05 *** 0.01. Robust standard error clustered at village level. Each column in each Panel represents a separate linear probability regression. All regressions include village fixed effects, province-specific age FEs, Han-specific age FEs, and control for household size. Columns 2 to 5 sequentially drop villages with more than 90, 70, 50, and 10 percent Han population. Samples for columns 5 and 6 include only minority and Han individuals, respectively. For these regressions, we consider individuals who responded to the written or spoken Mandarin ability questions.
}\\
\end{tabular}
\end{adjustbox}
\end{table}
\newpage

\subsection{Additional Mandarin ability regressions with non-respondents\label{sec:app:esti:mandwithna}}
In this section, we test the robustness of not including non-response to the language ability questions. In Table \ref{tab:interhan:mandwritetalkm2c:nsfto:withna}, we add non-respondents to the written and spoken Mandarin ability regressions from Table \ref{tab:interhan:mandwritetalkm2c:nsfto}. In Table \ref{tab:interhan:bygender:mandwritem2c:nsfto:withna}, we add non-respondents to the gender-specific written Mandarin ability regressions from Table \ref{tab:interhan:bygender:mandwritem2c:nsfto}.

As shown in Table \ref{tab:stats:indi:mandwritegroups}, non-respondents to the Mandarin ability questions, who account for 10 percent of the analytic sample, have substantially lower enrollment and attainment statistics compared to individuals who were reported by the household interviewee as having strong Mandarin ability. Given this, for the regressions in Table \ref{tab:interhan:mandwritetalkm2c:nsfto:withna} and \ref{tab:interhan:bygender:mandwritem2c:nsfto:withna}, we include non-respondents in the same group as individuals who have no, simple, or good Mandarin ability. We find that the results shown in Tables \ref{tab:interhan:bygender:mandwritem2c:nsfto:withna} and \ref{tab:interhan:bygender:mandwritem2c:nsfto} are largely invariant to the addition of non-respondents.

\begin{table}[htbp]
\centering
\caption{Regressions of Mandarin ability (including non-respondents) with school closure, Han-minority status interaction\label{tab:interhan:mandwritetalkm2c:nsfto:withna}}
\begin{adjustbox}{max width=1\textwidth}
\begin{tabular}{m{8.5cm} >{\centering\arraybackslash}m{1.5cm} >{\centering\arraybackslash}m{1.5cm} >{\centering\arraybackslash}m{1.5cm} >{\centering\arraybackslash}m{1.5cm} >{\centering\arraybackslash}m{1.5cm} >{\centering\arraybackslash}m{1.5cm}}
\toprule
&
\multicolumn{6}{c}{{\small Outcome:
strong
Mandarin
ability
=
1,
NA/no/simple/good
=
0}}
\\
\cmidrule(l{5pt}r{5pt}){2-7}
&
\multicolumn{1}{c}{\small
}
&
\multicolumn{5}{c}{
Village
Han
fraction
inclusion
threshold}
\\
\cmidrule(l{5pt}r{5pt}){2-2}
\cmidrule(l{5pt}r{5pt}){3-7}
&
\multicolumn{1}{C{1.5cm}}{All
villages}
&
\multicolumn{1}{C{1.5cm}}{Han
$<90$\%}
&
\multicolumn{1}{C{1.5cm}}{Han
$<70$\%}
&
\multicolumn{1}{C{1.5cm}}{Han
$<50$\%}
&
\multicolumn{1}{C{1.5cm}}{Han
$<10$\%}
&
\multicolumn{1}{C{1.5cm}}{Han
$\ge
90$\%}
\\
\midrule
\addlinespace[+1em]
\multicolumn{7}{L{19cm}}{\vspace*{1mm}\hspace*{-8mm}\textbf{Panel A: Written Mandarin ability}} \\
\multicolumn{7}{L{19cm}}{\vspace*{-3.5mm}\hspace*{-8mm}\textit{Baseline group: Child was 14-21 years old at village primary school closure year}} \\
\multicolumn{7}{L{19cm}}{\vspace*{-5mm}\hspace*{-8mm}{Child is minority interaction:}} \\
&                     &                     &                     &                     &                     &                     \\
\vspace*{0mm}\hspace*{5mm} Closure $\times$ age at closure 6-9&       -0.13\sym{***}&       -0.15\sym{***}&       -0.16\sym{***}&       -0.19\sym{***}&       -0.24\sym{***}&                     \\
                    &\vspace*{-2mm}{\footnotesize (0.041) }         &\vspace*{-2mm}{\footnotesize (0.042) }         &\vspace*{-2mm}{\footnotesize (0.045) }         &\vspace*{-2mm}{\footnotesize (0.046) }         &\vspace*{-2mm}{\footnotesize (0.055) }         &                     \\
\vspace*{0mm}\hspace*{5mm} Closure $\times$ age at closure 10-13&      -0.050\sym{*}  &      -0.063\sym{**} &      -0.065\sym{**} &      -0.073\sym{**} &      -0.076\sym{**} &                     \\
                    &\vspace*{-2mm}{\footnotesize (0.030) }         &\vspace*{-2mm}{\footnotesize (0.030) }         &\vspace*{-2mm}{\footnotesize (0.030) }         &\vspace*{-2mm}{\footnotesize (0.032) }         &\vspace*{-2mm}{\footnotesize (0.038) }         &                     \\
\vspace*{0mm}\hspace*{5mm} Closure $\times$ age at closure 22-29&      -0.022         &      -0.025         &      -0.019         &      -0.025         &      -0.063         &                     \\
                    &\vspace*{-2mm}{\footnotesize (0.030) }         &\vspace*{-2mm}{\footnotesize (0.031) }         &\vspace*{-2mm}{\footnotesize (0.032) }         &\vspace*{-2mm}{\footnotesize (0.034) }         &\vspace*{-2mm}{\footnotesize (0.044) }         &                     \\
\multicolumn{7}{L{19cm}}{\vspace*{-5mm}\hspace*{-8mm}{Child is Han interaction:}} \\&                     &                     &                     &                     &                     &                     \\
\vspace*{0mm}\hspace*{5mm} Closure $\times$ age at closure 6-9&       0.067         &      -0.049         &      -0.045         &      -0.030         &                     &        0.18\sym{**} \\
                    &\vspace*{-2mm}{\footnotesize (0.056) }         &\vspace*{-2mm}{\footnotesize (0.075) }         &\vspace*{-2mm}{\footnotesize (0.099) }         &\vspace*{-2mm}{\footnotesize (0.14) }         &                     &\vspace*{-2mm}{\footnotesize (0.075) }         \\
\vspace*{0mm}\hspace*{5mm} Closure $\times$ age at closure 10-13&      -0.038         &      -0.022         &     -0.0019         &      -0.022         &                     &      -0.043         \\
                    &\vspace*{-2mm}{\footnotesize (0.044) }         &\vspace*{-2mm}{\footnotesize (0.061) }         &\vspace*{-2mm}{\footnotesize (0.092) }         &\vspace*{-2mm}{\footnotesize (0.12) }         &                     &\vspace*{-2mm}{\footnotesize (0.060) }         \\
\vspace*{0mm}\hspace*{5mm} Closure $\times$ age at closure 22-29&      -0.031         &      0.0029         &      -0.068         &      -0.095         &                     &      -0.031         \\
                    &\vspace*{-2mm}{\footnotesize (0.037) }         &\vspace*{-2mm}{\footnotesize (0.043) }         &\vspace*{-2mm}{\footnotesize (0.047) }         &\vspace*{-2mm}{\footnotesize (0.063) }         &                     &\vspace*{-2mm}{\footnotesize (0.057) }         \\
\midrule
\hspace*{3mm}Observations        &       11637         &        9197         &        8371         &        7653         &        5686         &        2451         \\

\midrule
\addlinespace[+1em]
\multicolumn{7}{L{19cm}}{\vspace*{1mm}\hspace*{-8mm}\textbf{Panel B: Spoken Mandarin ability}} \\
\multicolumn{7}{L{19cm}}{\vspace*{-3.5mm}\hspace*{-8mm}\textit{Baseline group: Child was 14-21 years old at village primary school closure year}} \\
\multicolumn{7}{L{19cm}}{\vspace*{-5mm}\hspace*{-8mm}{Child is minority interaction:}} \\
&                     &                     &                     &                     &                     &                     \\
\vspace*{0mm}\hspace*{5mm} Closure $\times$ age at closure 6-9&      -0.030         &      -0.030         &      -0.032         &      -0.046         &      -0.082         &                     \\
                    &\vspace*{-2mm}{\footnotesize (0.046) }         &\vspace*{-2mm}{\footnotesize (0.048) }         &\vspace*{-2mm}{\footnotesize (0.050) }         &\vspace*{-2mm}{\footnotesize (0.052) }         &\vspace*{-2mm}{\footnotesize (0.070) }         &                     \\
\vspace*{0mm}\hspace*{5mm} Closure $\times$ age at closure 10-13&      -0.016         &      -0.026         &      -0.025         &      -0.028         &      -0.030         &                     \\
                    &\vspace*{-2mm}{\footnotesize (0.032) }         &\vspace*{-2mm}{\footnotesize (0.033) }         &\vspace*{-2mm}{\footnotesize (0.034) }         &\vspace*{-2mm}{\footnotesize (0.036) }         &\vspace*{-2mm}{\footnotesize (0.048) }         &                     \\
\vspace*{0mm}\hspace*{5mm} Closure $\times$ age at closure 22-29&      -0.019         &      -0.015         &      -0.013         &      -0.018         &      -0.064         &                     \\
                    &\vspace*{-2mm}{\footnotesize (0.029) }         &\vspace*{-2mm}{\footnotesize (0.029) }         &\vspace*{-2mm}{\footnotesize (0.030) }         &\vspace*{-2mm}{\footnotesize (0.031) }         &\vspace*{-2mm}{\footnotesize (0.040) }         &                     \\
\multicolumn{7}{L{19cm}}{\vspace*{-5mm}\hspace*{-8mm}{Child is Han interaction:}} \\&                     &                     &                     &                     &                     &                     \\
\vspace*{0mm}\hspace*{5mm} Closure $\times$ age at closure 6-9&       0.052         &      -0.047         &       0.031         &      -0.051         &                     &        0.11         \\
                    &\vspace*{-2mm}{\footnotesize (0.052) }         &\vspace*{-2mm}{\footnotesize (0.076) }         &\vspace*{-2mm}{\footnotesize (0.070) }         &\vspace*{-2mm}{\footnotesize (0.094) }         &                     &\vspace*{-2mm}{\footnotesize (0.071) }         \\
\vspace*{0mm}\hspace*{5mm} Closure $\times$ age at closure 10-13&     -0.0040         &      -0.034         &       0.079         &      0.0051         &                     &      0.0065         \\
                    &\vspace*{-2mm}{\footnotesize (0.036) }         &\vspace*{-2mm}{\footnotesize (0.058) }         &\vspace*{-2mm}{\footnotesize (0.068) }         &\vspace*{-2mm}{\footnotesize (0.099) }         &                     &\vspace*{-2mm}{\footnotesize (0.051) }         \\
\vspace*{0mm}\hspace*{5mm} Closure $\times$ age at closure 22-29&      -0.014         &       0.016         &      -0.017         &      -0.027         &                     &     -0.0075         \\
                    &\vspace*{-2mm}{\footnotesize (0.039) }         &\vspace*{-2mm}{\footnotesize (0.053) }         &\vspace*{-2mm}{\footnotesize (0.066) }         &\vspace*{-2mm}{\footnotesize (0.086) }         &                     &\vspace*{-2mm}{\footnotesize (0.051) }         \\
\midrule
\hspace*{3mm}Observations        &       11637         &        9197         &        8371         &        7653         &        5686         &        2451         \\
\bottomrule
\addlinespace[-1.5em]
\multicolumn{7}{L{19.7cm}}{
\footnotesize\justify
Statistical significance:* 0.10 ** 0.05 *** 0.01. Robust standard error clustered at village level. Each column in each Panel represents a separate linear probability regression. All regressions include village fixed effects, province-specific age FEs, Han-specific age FEs, and control for household size. Columns 2 to 5 sequentially drop villages with more than 90, 70, 50, and 10 percent Han population. Samples for columns 5 and 6 include only minority and Han individuals, respectively. For these regressions, we group individuals who did not respond to the written or spoken Mandarin ability questions along with individuals who responded as having no, simple, or good Mandarin ability. See Appendix Table \ref{tab:stats:indi:mandwritegroups} for how enrollment and attainment statistics for non-respondents differ from respondents.
}\\
\end{tabular}
\end{adjustbox}
\end{table}
\begin{table}[htbp]
\centering
\caption{Regressions of written Mandarin ability (including non-respondents) with school closure, Han-minority status interaction by gender\label{tab:interhan:bygender:mandwritem2c:nsfto:withna}}
\begin{adjustbox}{max width=1\textwidth}
\begin{tabular}{m{8.5cm} >{\centering\arraybackslash}m{1.5cm} >{\centering\arraybackslash}m{1.5cm} >{\centering\arraybackslash}m{1.5cm} >{\centering\arraybackslash}m{1.5cm} >{\centering\arraybackslash}m{1.5cm} >{\centering\arraybackslash}m{1.5cm}}
\toprule
&
\multicolumn{6}{c}{{\small Outcome:
strong
written
Mandarin
ability
=
1,
NA/no/simple/good
=
0}}
\\
\cmidrule(l{5pt}r{5pt}){2-7}
&
\multicolumn{3}{c}{\textbf{
Female}}
&
\multicolumn{3}{c}{\textbf{
Male}}
\\
\cmidrule(l{5pt}r{5pt}){2-4}
\cmidrule(l{5pt}r{5pt}){5-7}
&
\multicolumn{1}{C{1.5cm}}{All
villages}
&
\multicolumn{1}{C{1.5cm}}{Han
$<90$\%
villages}
&
\multicolumn{1}{C{1.5cm}}{Han
$<50$\%
villages}
&
\multicolumn{1}{C{1.5cm}}{All
villages}
&
\multicolumn{1}{C{1.5cm}}{Han
$<90$\%
villages}
&
\multicolumn{1}{C{1.5cm}}{Han
$<50$\%
villages}
\\
\midrule
\multicolumn{7}{L{19cm}}{\vspace*{-3.5mm}\hspace*{-8mm}\textit{Baseline group: Child was 14-21 years old at village primary school closure year}} \\            \multicolumn{7}{L{19cm}}{\vspace*{-5mm}\hspace*{-8mm}{Child is minority interaction:}} \\&                     &                     &                     &                     &                     &                     \\
\vspace*{0mm}\hspace*{5mm} Closure $\times$ age at closure 6-9&       -0.14\sym{**} &       -0.13\sym{**} &       -0.15\sym{***}&       -0.11\sym{**} &       -0.13\sym{**} &       -0.18\sym{***}\\
                    &\vspace*{-2mm}{\footnotesize (0.053) }         &\vspace*{-2mm}{\footnotesize (0.054) }         &\vspace*{-2mm}{\footnotesize (0.057) }         &\vspace*{-2mm}{\footnotesize (0.051) }         &\vspace*{-2mm}{\footnotesize (0.053) }         &\vspace*{-2mm}{\footnotesize (0.054) }         \\
\vspace*{0mm}\hspace*{5mm} Closure $\times$ age at closure 10-13&       -0.10\sym{**} &       -0.12\sym{***}&       -0.12\sym{**} &      -0.012         &      -0.025         &      -0.044         \\
                    &\vspace*{-2mm}{\footnotesize (0.044) }         &\vspace*{-2mm}{\footnotesize (0.044) }         &\vspace*{-2mm}{\footnotesize (0.047) }         &\vspace*{-2mm}{\footnotesize (0.041) }         &\vspace*{-2mm}{\footnotesize (0.042) }         &\vspace*{-2mm}{\footnotesize (0.046) }         \\
\vspace*{0mm}\hspace*{5mm} Closure $\times$ age at closure 22-29&      -0.011         &     -0.0042         &      -0.033         &      -0.021         &      -0.032         &      -0.023         \\
                    &\vspace*{-2mm}{\footnotesize (0.036) }         &\vspace*{-2mm}{\footnotesize (0.035) }         &\vspace*{-2mm}{\footnotesize (0.038) }         &\vspace*{-2mm}{\footnotesize (0.044) }         &\vspace*{-2mm}{\footnotesize (0.046) }         &\vspace*{-2mm}{\footnotesize (0.048) }         \\
\multicolumn{7}{L{19cm}}{\vspace*{-5mm}\hspace*{-8mm}{Child is Han interaction:}} \\&                     &                     &                     &                     &                     &                     \\
\vspace*{0mm}\hspace*{5mm} Closure $\times$ age at closure 6-9&       0.023         &      -0.070         &                     &        0.12         &      -0.018         &                     \\
                    &\vspace*{-2mm}{\footnotesize (0.071) }         &\vspace*{-2mm}{\footnotesize (0.11) }         &                     &\vspace*{-2mm}{\footnotesize (0.077) }         &\vspace*{-2mm}{\footnotesize (0.093) }         &                     \\
\vspace*{0mm}\hspace*{5mm} Closure $\times$ age at closure 10-13&      -0.080         &      -0.079         &                     &     -0.0077         &       0.023         &                     \\
                    &\vspace*{-2mm}{\footnotesize (0.059) }         &\vspace*{-2mm}{\footnotesize (0.080) }         &                     &\vspace*{-2mm}{\footnotesize (0.056) }         &\vspace*{-2mm}{\footnotesize (0.071) }         &                     \\
\vspace*{0mm}\hspace*{5mm} Closure $\times$ age at closure 22-29&      -0.013         &     -0.0092         &                     &      -0.046         &     -0.0040         &                     \\
                    &\vspace*{-2mm}{\footnotesize (0.049) }         &\vspace*{-2mm}{\footnotesize (0.064) }         &                     &\vspace*{-2mm}{\footnotesize (0.053) }         &\vspace*{-2mm}{\footnotesize (0.062) }         &                     \\
\midrule
\hspace*{3mm}Observations        &        5476         &        4328         &        3328         &        6161         &        4869         &        3830         \\
\bottomrule
\addlinespace[-1.5em]
\multicolumn{7}{L{19.7cm}}{
\footnotesize\justify
Statistical significance:* 0.10 ** 0.05 *** 0.01. Robust standard error clustered at village level. Each column represents a separate linear probability regression. All regressions include village fixed effects, province-specific age FEs, Han-specific age FEs, and control for household size. Columns 1, 2, and 3 regressions include only females. Columns 4, 5, and 6 regressions include only males. Columns 2 to 3 (and 5 to 6) sequentially drop villages with more than 90 and 50 percent Han population. Samples for columns 3 and 6 regressions include only minority individuals. For these regressions, we group individuals who did not respond to the written or spoken Mandarin ability questions along with individuals who responded as having no, simple or good Mandarin ability. See Appendix Table \ref{tab:stats:indi:mandwritegroups} for how enrollment and attainment statistics for non-respondents differ from respondents.
}\\
\end{tabular}
\end{adjustbox}
\end{table}
\clearpage
\pagebreak

\end{document}